%% file: master.tex
\documentclass[10pt]{iopart}

\usepackage{iopams} 

\usepackage[utf8]{inputenc}
\usepackage[english]{babel}
\usepackage[T1]{fontenc}
\usepackage{graphicx}
\usepackage{color}
\usepackage{enumitem}
\usepackage{harvard}
\begin{document}

\title[Accretion heating and thermal conduction in protoplanetary disks]{The impact of accretion heating and thermal conduction on the dead zone of protoplanetary disks}

\author{B. N. Schobert, A. G. Peeters and F. Rath}

\address{Physics Department, University of Bayreuth, Universit\"atsstra{\ss}e 30 Bayreuth, Germany}
\ead{benjamin.schobert@uni-bayreuth.de}
\vspace{10pt}
\begin{indented}
\item[]April 2019
\end{indented}

\begin{abstract}
The paper investigates the influence of accretion heating and turbulent heat conduction on the equilibrium of protoplanetary disks, extending the 2D axis-symmetric passive disk model of Flock (Flock et al. 2016, ApJ 827, 144). The model includes dust sublimation and radiative transfer with the flux-limited diffusion approximation, and predicts the density and temperature profiles as well as the dust to gas ratio of the disk. It is shown that the accretion heating can have a large impact:  For accretion rates above $5\cdot 10^{-8}\,M_\odot/$yr a zone forms behind the silicate condensation front with sufficiently high temperature to sublimate the dust and form a gaseous cavity. Assuming a Prandtl number  $\sim 0.7$, it is furthermore shown that the turbulent heat conduction cannot be neglected in the evaluation of the temperature profile. While the inner rim position is not affected by viscous heating, the dead zone edge shifts radially outward for higher accretion rates.
\end{abstract}

\vspace{2pc}
\ioptwocol

\section{Introduction}

The observational techniques used to picture exoplanets have constantly been enhanced culminating in the recent observation of a nascent exoplanet, PDS 70b \cite{AM,MK}. In order to explain the formation of planets it is paramount to understand the physics of the protoplanetary disks where they originate. Modern terrestrial and satellite telescopes use large parts of the electromagnetic spectrum, especially near infra-red (NIR), to resolve young stellar objects and their disks. Thus they deliver valuable data for the models to be tested against or build upon. Of special interest are Herbig Ae/Be stars, since they are slightly more massive (2-5 $M_\odot$) but also younger than the sun \cite{Her}. Interferometric examinations of such stars allow for conclusions about their disks, however the interpretation has proven difficult.

Therefore the theoretical model of such disks has been continually augmented.
Early contributions are the $\alpha$-prescription to estimate the turbulent viscosity \cite{SS} and the flaring disk model \cite{CG} for passive disks. In passive disks the effects of viscous dissipation are neglected which is assumed to be a good approximation for accretion rates below $2 \cdot 10^{-7}\,M_\odot$/yr \cite{vdA04}. The effect of viscous dissipation on the vertical structure of passive disks has been treated in 1D for T Tauri stars by \cite{DA}.

Because rocky planets originate in the inner regions of the disk, this is a focus of research \cite{KLGT}. The temperatures close to the star are sufficient to sublimate the dust and an inner gaseous zone forms.  In \cite{DDN} this area was modelled to be limited by a cylindrical sublimation front.
The stellar radiation is strongly absorbed at this front, leading to a region of lower temperature behind the front known as the dead zone. In this zone the ionization is expected to be insufficient for MRI \cite{BH} and, consequently, also the turbulent viscosity is possibly small \cite{Tur}.

The geometry of the inner rim was further explored by \cite{FL16} for Herbig Ae/Be stars. Using a 2D radiative transfer model based on the flux-limited diffusion approximation a rounded off rim was obtained with four distinct regions: an inner gaseous hole, an optically thin dust halo, the round irradiated rim and a shadowed region behind it. That work also included accretion heating through viscous dissipation in some models with rates up to $\dot{M}=10^{-8}\,M_\odot/$yr. To achieve this a hydrostatic simulation was run first and then the results were used as initial conditions for a second hydrodynamic simulation based on previous work \cite{FL13}. At this accretion rate a small increase in temperature of only 10\% in the shadowed region was found. The density and temperature structure remained similar to that of a hydrostatic model without accretion heating.

In this paper the above mentioned model is extended to include accretion heating through viscous dissipation at rates beyond $10^{-8}\,M_\odot/$yr. Indeed, experimental data suggest that accretion rates up to $10^{-6} \, M_\odot/\textrm{yr}$ are common \cite{vdA04} and that 25\% of HAe stars have $\dot{M} > 10^{-7} \, M_\odot/\textrm{yr}$ \cite{LNT}. Furthermore, the accretion rate of a star decreases with time and is therefore likely to be high in the early stages of development that are relevant for the planet formation. The accretion heating is modelled within a single hydrostatic simulation by modifying the energy balance to include viscous heating. The model is also expanded by adding thermal conduction, so that turbulent heat transport is no longer neglected. This is a reasonable addition considering that turbulence plays a vital role in momentum transfer, it might impact heat transfer as well. The term is estimated by assuming a Prandtl number of order unity and turns out to be relevant as well, especially for cases of high viscous heating. 
Therefore it needs to be treated simultaneously, as will be discussed later in section \ref{sec:tc}.

This paper is structured as follows: section 2 explains the model equations needed to describe the disk. In section 3 the numerical model used to solve the equations is presented. Additionally, the boundary conditions are motivated and a benchmark to validate the implementation is performed. Section 4 outlines the results and the qualitative changes compared to a model without accretion heating and thermal conduction. Section 5 discusses the results and possible limitations of the work. The paper concludes in section 6 with a summary of the results and a short outlook.

\section{Model equations}

The model used in this paper closely follows the model of \cite{FL16}, extending the latter to include viscous heating and turbulent heat conduction. In this section the model equations are presented. Since there is a considerable overlap with \cite{FL16} at some points the discussion is brief and the reader is referred to the latter reference.
The model describes the density and temperature distributions of the disk for a star with known parameters, specifically mass $M_*$, radius $R_*$ and luminosity $L_*$ or equivalently surface temperature $T_*$.

The model relies on the large difference in relevant timescales. These are the dynamical timescale in which hydrostatic equilibrium is achieved $t_{\textrm{\footnotesize dyn}}$, typically the duration of one orbit, the timescale of radiative transfer in the optically thick regime $t_{\textrm{\footnotesize rad}}^{\tau \gg 1}$, typically 50 orbits, and the viscous timescale $t_{\textrm{\footnotesize visc}}$ over which the surface density changes, typically $10^4$ orbits. For typical disk parameters the ordering is
\begin{equation}
	t_{\textrm{\footnotesize dyn}} \ll t_{\textrm{\footnotesize rad}}^{\tau \gg 1} \ll t_{\textrm{\footnotesize visc}}.
\end{equation}
The timescale followed in the solution is that of the surface density, it changes slowly
over many iterations of the temperature profile and is the last variable to converge.
The temperature profile for each step is obtained by solving the equations for the internal and radiation energy, which operate on the intermediary timescale.
For the density one can assume vertical hydrostatic balance for each step because that forms on a even shorter timescale
\cite{FL16}.

\subsection{Hydrostatic equilibrium}\label{he}

The density structure is calculated assuming a hydrostatic equilibrium between ideal gas pressure, the centrifugal force and gravity. It is convenient to use spherical coordinates $(r, \theta, \phi)$ for the model problem, since they facilitate a straightforward integration of the radiation from the star along its optical path. The polar axis of the coordinate system is oriented along the rotational axis of the star, so that the protoplanetary disk lies in the equatorial plane. This means the azimuthal velocity of the gas will be significantly greater than radial or polar velocities $v_{\phi} \gg v_r, v_\theta$. Neglecting the latter the hydrostatic equations are
\begin{equation}
\frac{\partial p}{\partial r} = -\rho \frac{\partial \Phi}{\partial r} + \frac{\rho \, v_\phi^2}{r} \label{HE1}
\end{equation}
\begin{equation}
\frac{1}{r} \frac{\partial p}{\partial \theta} = \frac{\rho \, v_\phi^2}{r \tan \theta} \, , \label{HE2}
\end{equation}
where $\rho$ is the gas density, $v_\phi$ is the velocity in azimuthal direction, $\Phi = G M_*/r$ is the gravitational potential and $p$ is the pressure. For closure with the thermodynamic equations an ideal gas is assumed
\begin{equation}
p = \frac{\rho k_B T}{\mu_g u}, \label{pres}
\end{equation}
with temperature $T$, Boltzmann constant $k_B$, mean molecular weight $\mu_g$ and atomic mass unit $u$.

\subsection{Radiative Hydrodynamics}\label{rh}

The temperature distribution is the solution of a coupled system of equations for the radiation energy density $E_R$ and the internal energy density of the gas $\epsilon$. These equations represent a subset of ideal radiative magnetohydrodynamics (RMHD) with magnetic and electric field neglected. The two coupled equations for the radiation and internal energy are
\begin{equation}
	\partial_t \rho \epsilon = -\sigma c (a_R T^4 - E_R) - \nabla \cdot \mathbf{F_*} \label{RHD1}
\end{equation}
\begin{equation}
	\partial_t E_R -\nabla \frac{c \lambda}{\sigma}\nabla E_R = +\sigma c (a_R T^4 - E_R),
\end{equation}
where $\sigma$ is the mean opacity, $a_R = 4 \sigma_B/c$ is the radiation constant with $\sigma_B$ being the Stefan-Boltzmann constant, $\mathbf{F_*}$ is the irradiation flux from the star, $c$ is the vacuum speed of light and $\lambda$ is the flux limiter. The flux limiter acts as a diffusion constant for the radiation energy density, it is taken from \cite{LP} and has the following form
\begin{equation}
	\lambda = \frac{2+R}{6+3R+R^2} \qquad \textrm{with}
\end{equation}
\begin{equation}
	R = \frac{|\nabla E_R|}{\sigma E_R},
\end{equation}
which fulfils $\lim\limits_{R \rightarrow 0}{\lambda(R)}=1/3$ in the optically thick limit and $\lim\limits_{R \rightarrow \infty}{\lambda(R)}=0$ in the optically thin limit. Since the product $R \lambda(R)$ can never exceed unity, this flux-limited diffusion theory (FDT) preserves causality by never allowing the radiative flux to exceed the radiation energy density times the speed of light in vacuum.

For closure between internal energy and temperature again the ideal gas approximation
\begin{equation}
	\rho \epsilon = c_V \rho T
\end{equation}
is used. Here $c_V$ is the specific heat capacity. This yields for (\ref{RHD1})
\begin{equation}
	c_V \partial_t \rho T = -\sigma c (a_R T^4 - E_R) - \nabla \cdot \mathbf{F_*} \, .
\end{equation}
For the irradiation flux black body radiation times an attenuation factor is assumed
\begin{equation}
	\mathbf{F_*}(r) = \left( \frac{R_*}{r} \right)^2 \sigma_B T_*^4 e^{-\tau_*},
\end{equation}
where $\tau$ is the optical depth
\begin{equation}
	\tau_* = \int\limits_{R_*}^{r}\sigma_* dr = \tau_0 + \int\limits_{r_{\mathrm{min}}}^{r}\sigma_* dr \label{optd}
\end{equation}
and	$\tau_0 = \kappa_{\mathrm{gas}} \rho(r_{\mathrm{min}})(r_{\mathrm{min}}-3R_*)$.

Furthermore, the mean opacities at the typical wavelengths of the stellar light and the rim's thermal emission are defined as
\begin{equation}
	\sigma_* = \rho_{\mathrm{dust}}\kappa_{\mathrm{dust}}(\nu_*)+ \rho_{\mathrm{gas}}\kappa_{\mathrm{gas}}
\end{equation}
\begin{equation}
	\sigma = \rho_{\mathrm{dust}}\kappa_{\mathrm{dust}}(\nu_{\mathrm{rim}})+ \rho_{\mathrm{gas}}\kappa_{\mathrm{gas}}
\end{equation}
\begin{equation}
	\rho_{\mathrm{dust}} = f_{\mathrm{d2g}} \rho_{\mathrm{gas}}  \, .
\end{equation}
Here $\kappa_{\mathrm{gas}}$ is the frequency averaged opacity of the gas. 
Finally $f_{\mathrm{d2g}}$ is the dust-to-gas ratio of the respective densities and its calculation is explained in the following section.

\subsection{Dust sublimation and opacities}

The most crucial effect for the evolution of the disk is arguably the dust sublimation. Because the dust absorbs more radiation than the gas, it is strongly heated by the star and through its own so-called back-warming, i.e. the infrared radiation emitted by the dust. A small amount of dust can substantially decrease the radiation that reaches the area behind it, making the transition between vapour and condensed dust very thin.

In order to resolve this thin layer the dust sublimation formula from \cite{FL16} smooths the transition over a temperature range of 100 K and uses the tangens hyperbolicus as a model function. The formula is
\begin{equation*}
	f_{\mathrm{d2g}} =  \frac{f_{\Delta \tau}}{2} \left\lbrace 1- \tanh \left[\left(\frac{T-T_{\mathrm{ev}}}{100\, \textrm{K}}\right)^3\right] \right\rbrace \cdot
\end{equation*}
\begin{equation*}
	\qquad \quad \left\lbrace \frac{1- \tanh(1-\tau_*)}{2} \right\rbrace \qquad \textrm{if} \enspace T>T_{\mathrm{ev}}
\end{equation*}
\begin{equation}
	= \frac{f_0}{2} \left\lbrace 1- \tanh (20-\tau_*)\right\rbrace + f_{\Delta \tau} \quad \, \textrm{if} \enspace T<T_{\mathrm{ev}} \, \textrm{,} \label{d2g}
\end{equation}
with the dust evaporation temperature $T_{\mathrm{ev}}$, the reference dust-to-gas ratio $f_0$ and the transition dust-to-gas ratio $f_{\Delta \tau}$.
For the dust evaporation temperature the fitting model proposed by \cite{IN}
\begin{equation}
	T_{\mathrm{ev}} = 2000 \, \textrm{K} \, \left(\frac{\rho}{1 \, \textrm{g\,cm}^{-3}}\right)^{0.0195}
\end{equation}
is used. It describes the dependence of the evaporation temperature on the gas density for silicate grains.
The transition dust-to-gas ratio $f_{\Delta \tau}$ is defined as
\begin{equation}
	f_{\Delta \tau} = \frac{\Delta \tau_*}{\rho_{\mathrm{gas}} \kappa_{\mathrm{dust}}(\nu_*) \Delta r}= \frac{0.3}{\rho_{\mathrm{gas}} \kappa_{\mathrm{dust}}(\nu_*) \Delta r}
\end{equation}
with $\Delta r$ being the radial size of one grid cell. The transition optical depth of $\Delta \tau_* = 0.3$ is chosen so that the absorption of the radiation at the rim can be resolved \cite{FL16}.
Furthermore, it is useful for numeric stability to impose a minimum value of $f_{\mathrm{d2g}}^{\mathrm{min}}= 10^{-10}$. The maximum value of $f_0 = 10^{-2}$ is chosen, because it reflects the amount of dust present in the interstellar medium \cite{LiDrain} and therefore represents the maximum ratio in the protoplanetary disk.

The dust sublimation formula describes gradual building up of the dust halo for temperatures lower than $T_{\mathrm{ev}}$ and the actual sublimation front for temperatures above. By design, the formula for the dust halo has an upper limit $f_{\Delta \tau}$, which is reached for optical depths larger than one, close to the evaporation temperature. Beyond the condensation front the dust-to-gas ratio grows with the optical depth and reaches its maximum for $\tau_* \ge 20$.

Similar to \cite{FL16} an average value of $\kappa_{\mathrm{gas}} = \mathbf{10^{-4} \, \textrm{cm}^2 \textrm{g}^{-1}}$ is used in the computations of this paper. This value ensures that the optical depth $\tau_*$ remains small enough inside the gaseous inner disk, preventing the absorption of too much stellar radiation in this region, which would result in the inner rim moving too close to the star. 

For the opacity of the dust two wavelengths are important, the stellar light's and the thermal radiation of the rim at the condensation temperature. In the specific case of a star with a surface temperature ${T_* = 10,000\,\textrm{K}}$ a dust opacity of ${\kappa_{\mathrm{dust}}(\nu_*) = \mathbf{2100 \, \textrm{cm}^2 \textrm{g}^{-1}}}$ is calculated in \cite{FL16} using the MieX code by \cite{WV}. Using the same method an opacity of ${\kappa_{\mathrm{dust}}(\nu_{\mathrm{rim}}) = \mathbf{700 \, \textrm{cm}^2 \textrm{g}^{-1}}}$ is obtained for thermal radiation of approximately 1300 K, which represents a typical dust sublimation temperature.

\subsection{Surface density}

The surface density $\Sigma$ is modelled using the steady thin disk approximation \cite{CC}
\begin{equation}
	\Sigma = \frac{\dot{M}}{3\pi \nu_t }\left[ 1- \left(\frac{R_*}{r}\right)^{0.5}\right] \approx \frac{\dot{M}}{3\pi \nu_t }\label{sd},
\end{equation}
where the latter approximation applies at distances much larger than the radius of the star. Here $\dot{M}$ is the accretion rate, $r$ the radial distance and $\nu_t$ the kinematic turbulent viscosity.

As an estimate for the viscosity the $\alpha$-viscosity prescription introduced by \cite{SS} is employed:
\begin{equation}
	\nu_t = \alpha H c_s = \frac{\alpha c_s^2}{\Omega} \, ,
\end{equation}
where the pressure scale height $H= c_s/ \Omega$ was used to expresses the kinematic turbulent viscosity $\nu_t$ through the local speed of sound $c_s = \sqrt{\partial p/ \partial \rho}$ and the Kepler rotation frequency $\Omega = \sqrt{GM_*/R^3}$. The constant $\alpha$ is in the order of $10^{-2}$ for turbulent flow.

With a given viscosity it is possible to calculate the surface density and the effect of the viscous dissipation on the temperature.

\subsection{Viscous dissipation}

The significant turbulent viscosity suggests that heat dissipation needs consideration in the energy balance, especially for active disks. In contrast to \cite{FL16} this effect is included in this paper by introducing the heating term in the equation for the evolution of the internal energy.
The viscous heating term can be obtained from the Navier-Stokes equation by taking the scalar product with the velocity field. This yields an additional term $Q_{\mathrm{heat}}$ in (\ref{RHD1}):
\begin{equation}
\partial_t \rho \epsilon = -\sigma c (a_R T^4 - E_R) - \nabla \cdot \mathbf{F_*} + Q_{\mathrm{heat}},
\end{equation}
with
\begin{equation}
	Q_{\mathrm{heat}} =  \boldsymbol{\sigma} \colon \nabla \mathbf{v}, \label{eq:Q_heat}
\end{equation}
where $\colon$ denotes a double contraction. In spherical coordinates assuming a dominant azimuthal velocity the expression of (\ref{eq:Q_heat}) simplifies to 
\begin{equation}
	Q_{\mathrm{heat}} = \rho \nu_t \left[r \, \partial_r \, \Omega\right]^2.
\end{equation}
The effect of this new term is discussed in section \ref{viscHeat}.

\subsection{Thermal conduction}\label{sec:tc}

Since momentum eddy diffusivity and heat transfer eddy diffusivity are linked through the turbulent Prandtl number $Pr_t= c_p \nu_t \rho/k_t = \mathcal{O}(1)$,
the effect of thermal conduction should have a similar impact on the internal energy as the effect of viscous dissipation and therefore will also be considered. This new addition to the model will be shown to be important for the higher accretion rates studied in this paper.  Here $k_t$ is the turbulent thermal conductivity and $c_p$ is the specific heat at constant pressure. This yields for $k_t$
\begin{equation}
k_t = \frac{\rho \nu_t \Gamma c_V}{Pr_t},
\end{equation}
where $c_p= \Gamma c_v$, with $\Gamma$ being the adiabatic index. The addition of the turbulent heat conduction changes equation (\ref{RHD1}) to
\begin{equation}
	\partial_t \rho \epsilon = -\sigma c (a_R T^4 - E_R) - \nabla \cdot \mathbf{F_*} + Q_{\mathrm{heat}} + Q_{\mathrm{cond}}
\end{equation}
with
\begin{equation}
	Q_{\mathrm{cond}} =  k_t \nabla^2 T \, .
\end{equation}
The changes through thermal conduction will be addressed in section  \ref{viscHeat}.

\section{Numerical implementation}
This section details the numerical implementation of the above described model as well as the set of initial and boundary conditions. Especially the boundary conditions pose a delicate problem that needs to be explored tentatively, because of their influence on the final solution. Also a benchmark for a specific set of parameters is presented.
The implementation in this paper is partially derived from the model
of \cite{FL13}, but deviates when expedient.\\
The code used in this paper can be found at \mbox{bitbucket.org/astro\_bayreuth/radiation\_code}. It is written in Matlab/ Octave and implicitly solves the radiation hydrodynamic equations.

\subsection{Iterative procedure}

The procedure to determine the final disk structure is threefold.

The first step is to calculate the hydrostatic equilibrium for a given temperature distribution using the equations of section \ref{he} to determine the density distribution. For the density in the equatorial plane $\rho_0$ the surface density and the pressure scale height $H = [k_B T r^3/(GM_*\mu_g u)]^{0.5}$ are used:
\begin{equation}
	\rho_0 = \frac{\Sigma}{\sqrt{2\pi}H} .
\end{equation}
Then, from the midplane outwards, density and pressure are integrated. At this point one also calculates the optical depth with formula (\ref{optd}) and the corresponding radiation flux $F_*$.

The second step is to implicitly solve the two coupled equations for the temperature and radiation field from section \ref{rh}. This is done using the BiCSTAB solver first presented by \cite{VDV}. For preconditioning the incomplete LU-factorization is used and the convergence criterion is met if the reduction of the $L_2$ norm of the residual $||\mathbf{r}||_2 / ||\mathbf{r}_{\mathrm{init}}||_2$ is $< 10^{-5}$. This deviates from \cite{FL13} where the matrix is Jacobi preconditioned, but faster convergence is found with incomplete LU-factorization.

The final step is to calculate the dust-to-gas ratio with formula (\ref{d2g}). After these three steps, the process is repeated starting again with the hydrostatic equilibrium using the newly obtained temperature distribution. The iterations are continued until the relative change of temperature and density field  
\begin{equation*}
	\max\left(\textrm{max}\left(\frac{T^{n}-T^{n-1}}{T^{n-1}}\right),
	\textrm{max}\left(\frac{\rho^{n}-\rho^{n-1}}{\rho^{n-1}}\right)\right),
\end{equation*}
 which is used as the convergence criterion, reaches a steady state.

For stability it has been found that it is advantageous to introduce the dust-to-gas ratio logarithmically in the first 30 steps of the iteration. 
The calculation begins with no dust present and then the effect is slowly introduced. This was done in \cite{FL16} as well, but only for the first five iterations. It is necessary to introduce the dust slower because of the different grid spacing. The grid is equidistantly spaced in radial and polar direction, in contrast to \cite{FL16} where a logarithmically spaced radial grid is employed. Therefore, to compensate for the bigger grid cells in the inner region, the model needs more steps to equilibrate during the dust introduction. Furthermore, up to three different time steps are used, initially smaller ones $dt_1$ for $N_1$ iterations then bigger ones $dt_2$ for $N_2$ iterations and finally again smaller ones $dt_3$ for $N_3$ iterations. This way the code converges into a stable configuration with less iterations. The small steps $dt_1 = dt_3$ are chosen as $10^4$\,s or $0.0056$ inner orbits, while the bigger ones $dt_2$ are $10^{12}$\,s or 560 thousand inner orbits in the benchmark case and $10^7$\,s or 5.6 inner orbits in all later cases.

\subsection{Initial and boundary conditions}

As stated in the previous section there is initially no dust present and the dust-to-gas ratio is zero everywhere. Therefore the initial temperature distribution is that of an optically thin gas:
\begin{equation}
	T_{\mathrm{thin}}(r) = \left(\frac{1}{\epsilon_{\mathrm{gas}}}\right)^{0.25} \left(\frac{R_*}{2r}\right)^{0.5} T_*  \label{eq:T_thin}
\end{equation}
with $\epsilon_{\mathrm{gas}} = 1$. Using the initial temperature the surface density can be calculated using equation (\ref{sd}). (Note that for the benchmark presented below the density has been set constant.) 
Then the density can be integrated. The initial condition for the radiation energy density is
${E_R = a_R T^4}$
with the radiation constant $a_R$.

The boundary condition for the temperature field is zero gradient on all four edges. For the energy density the boundaries are fixed. At the outer edge at ${E_R = 3.4255 \cdot 10^{-6} \, \textrm{J}/\textrm{m}^3}$, which is equivalent to about 260 K. This is the temperature of the outer edge in \cite{FL16} and also close to the analytical expression in \cite{UOF}. At the inner edge the condition is
\begin{equation}
	E_R = (1-\e^{-\tau_0})a_R T_{\mathrm{thin}}^4(r_{\mathrm{min}})\label{inneredge},
\end{equation}
which accounts for the optical depth up to the inner boundary, and uses $\epsilon_{\mathrm{gas}}$ since there is no dust present in proximity to the star.

Along the upper and lower radial boundary the problem is more complex. While the gas is still optically thin, the boundary condition is given by inserting (\ref{eq:T_thin}) in (\ref{inneredge}) with a modified $\epsilon$ that includes the dust-to-gas ratio:
\begin{equation}
	\epsilon = \frac{\kappa_{\mathrm{gas}}+ f_{\mathrm{d2g}}\kappa_{\mathrm{dust}}(\nu_{\mathrm{rim}})}{\kappa_{\mathrm{gas}}+ f_{\mathrm{d2g}}\kappa_{\mathrm{dust}}(\nu_{*})} .
\end{equation}
This method is used up to the transition radius ${r_{\mathrm{trans}} = 0.83 \, \mathrm{AU}}$ that was calculated with expressions from \cite{UOF}, specifically equations (6), (17) and (19) therein. For these calculations a ratio between surface height and pressure scale height $z_*/H = 3.6$ is used. For the temperature in this region equation (21) from \cite{UOF} can be modified to
\begin{equation}
	T = \left\lbrace \frac{2}{\pi}\left[\arctan\left(\frac{r_\mathrm{trans}-r}{h}\right)+ \frac{\pi}{2}\right] \right\rbrace^{1/4} T_\mathrm{trans},
\end{equation}
where $T_\mathrm{trans}$ is the temperature at $r_\mathrm{trans}$ and $h = 0.04 \, r_\mathrm{trans}$.

This holds true until $T < T_\mathrm{flaring}$ with
\begin{equation}
T_\mathrm{flaring} = \frac{550 \, \textrm{K}}{r_{\mathrm{AU}}^{3/7}}  
\end{equation}
being the flaring disk approximation from \cite{CG}. The value of $550$\,K is also taken from \cite{FL16} and $r_{\mathrm{AU}}$ is the radius in astronomic units. Then the temperature $T_{\mathrm{flaring}}$ is used. All these temperatures are transformed into the corresponding radiation energy values analogue to (\ref{inneredge}) with the optical depth being capped at unity.

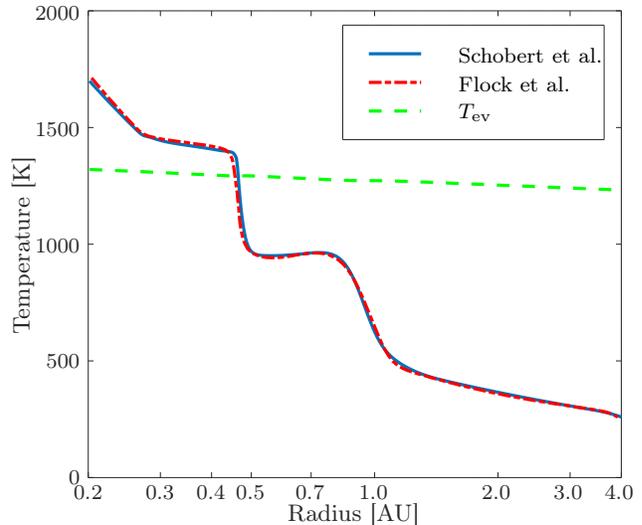
\begin{figure}
	\input{radialtemp.tex}
	\vspace{-1cm}
	\caption{Our midplane radial temperature profile (blue solid line) for model \texttt{S100} compared to the temperature profile of (Flock et al. 2016) (red dashed and dotted line). The evaporation temperature (green dashed line) is shown for reference.}
	\label{fig:benchTemp}
\end{figure}

\subsection{Benchmark}

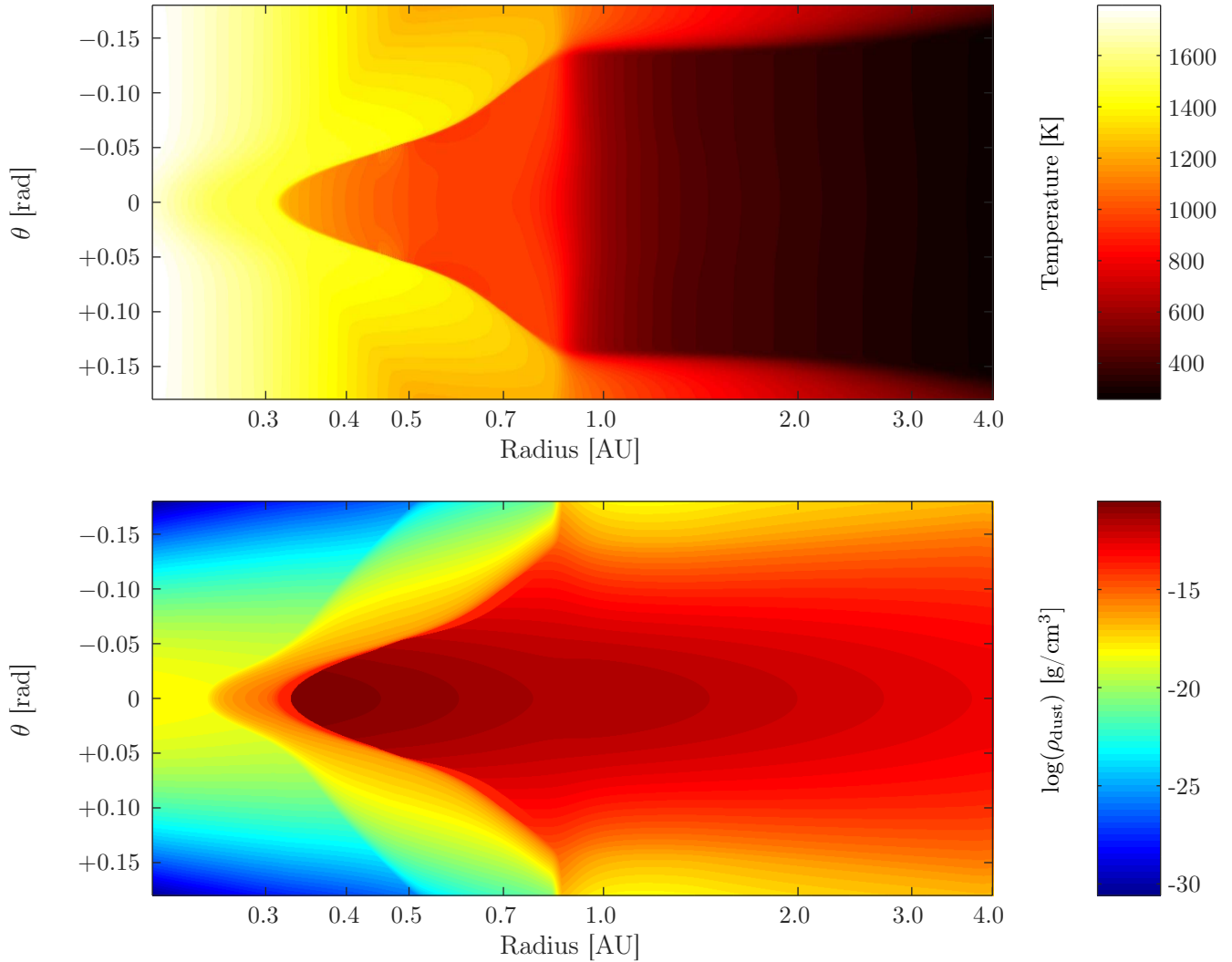
\begin{figure*}
	\input{temp.tex}
	\input{dustdens.tex}
	\vspace{-0.5cm}
	\caption{2D profiles of temperature (top) and the logarithm of the dust density in g/cm$^3$ to the base 10 (bottom) for the \texttt{jTC1e-7} case. The y-axis is the polar angle in rad offset by $\pi/2$ and the x-axis the radial distance in AU.}
	\label{fig:bench2d}
\end{figure*}
\begin{table}
	\caption{\label{tab1}Model \texttt{S100} setup parameters.}
	\begin{indented}
		\item[]\begin{tabular}{@{}llll}
			\br
			Parameter&Value\\
			\mr
			Surface density&100 g cm$^{-2}$, uniform\\
			$N_r \times N_\theta$&$1280 \times 129$\\
			$[r_{\mathrm{min}},\,r_{\mathrm{max}}]$&$[0.2\,\textrm{AU},\,4\,\textrm{AU}]$\\
			$[\theta_{\mathrm{min}},\,\theta_{\mathrm{max}}]$&$[\pi/2-0.18, \,\pi/2+0.18]$\\
			Stellar parameter&$T_* = 10 000$ K, $R_*=2.5 R_\odot$, $M_*=2.5M_\odot$\\
			Dust-to-gas ratio& $f_0=0.01$\\
			Time steps& $dt_1 = 10^4\,$s, $dt_2= 10^{12}\,$s\\
			Iterations& $N_1=40$, $N_2=20$\\
			\br
		\end{tabular}
	\end{indented}
\end{table}

To verify the compliance of the code with previous results a benchmark against the \texttt{S100} run from \cite{FL16} is presented. In this simulation both viscous heating as well as thermal conduction are neglected and the run uses a constant surface density in every iteration for comparability. This is the same setup as was used in the run which it is compared against. The parameters used are listed in table \ref{tab1}. They are typical for a Herbig Ae class star \cite{vdA}. The dust-to-gas ratio is fixed after the 40 small steps for faster convergence.

The resulting radial temperature can be seen in figure \ref{fig:benchTemp}. The temperatures are in very good agreement corroborating the correctness of the implementation.

Figure \ref{fig:bench2d} represents the 2D profile for a case with thermal conduction and without viscous heating. The parameters used are as listed in table \ref{tab:viscHeat} but with $N_3 = \mathbf{250}$.
It shows all the features, like the dust halo, a rounded off rim and a shadowed region behind the rim, that were also found in the \texttt{S100} run from \cite{FL16}. It follows, that the simulation code used in this paper reproduces both qualitatively as well as quantitatively the results of \cite{FL16}, giving confidence in the implementation. 

\section{Results}

This section presents the results for cases with viscous heating and thermal conduction, describes the influence of the accretion rate and analyses the effect on the structure of the rim.

\subsection{Rise in temperature through viscous dissipation and cooling through thermal conduction}\label{viscHeat}

\begin{table}
	\caption{\label{tab:viscHeat}Model setup parameters for figures \ref{fig:VHTemp} and \ref{fig:TCTemp}.}
	\begin{indented}
		\item[]\begin{tabular}{@{}llll}
			\br
			Parameter&Value\\
			\mr
			Surface density&$\Sigma(r)=\dot{M}/(3\pi\nu_t)$\\
			$N_r \times N_\theta$&$5120 \times 513$\\
			$[r_{\mathrm{min}},\,r_{\mathrm{max}}]$&$[0.2\,\textrm{AU},\,4\,\textrm{AU}]$\\
			$[\theta_{\mathrm{min}},\,\theta_{\mathrm{max}}]$&$[\pi/2-0.18, \,\pi/2+0.18]$\\
			Stellar parameter&$T_* = 10 000$ K, $R_*=2.5 R_\odot$, $M_*=2.5M_\odot$\\
			Dust-to-gas ratio& $f_0=0.01$\\
			Time steps& $dt_1 = 10^4\,$s, $dt_2= 10^{7}\,$s, $dt_3= 10^4\,$s\\
			Iterations& $N_1=30$, $N_2=40$, $N_3=50$\\
			$q_\tau$&0.1\\
			\br
		\end{tabular}
	\end{indented}
\end{table}

\begin{figure}
	\input{radialtemp2.tex}
	\vspace{-1cm}
	\caption{Comparison of four radial temperature profiles of runs including viscous heating (\texttt{VH1}) and/ or thermal conduction (\texttt{TC1}), the black solid line is for the model \texttt{TC0VH0}, the red dotted line is for \texttt{TC1VH0}, the blue dashed and dotted line is for \texttt{TC0VH1} and the purple solid line is for \texttt{TC1VH1}. The evaporation temperature (green dashed line) is shown for reference.}
	\label{fig:VHTemp}
\end{figure}
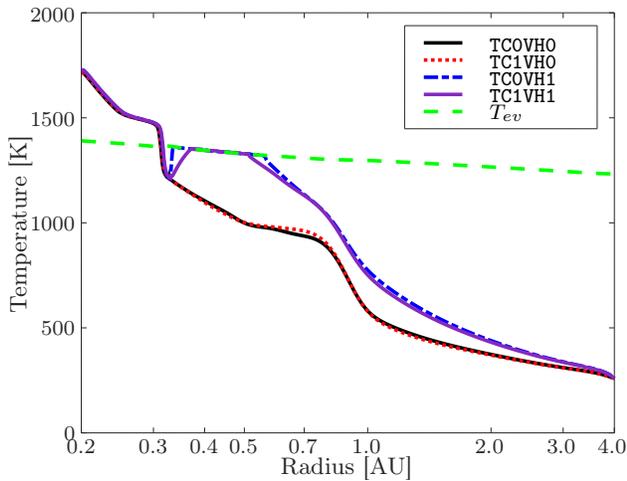

As a first step the influence of viscous heating and thermal conduction will be examined. To do so four cases are compared with viscous heating turned on (\texttt{VH1}) versus turned off (\texttt{VH0}) and thermal conduction enabled (\texttt{TC1}) or disabled (\texttt{TC0}) with otherwise identical parameters as listed in table \ref{tab:viscHeat}.

The accretion rate chosen is $5\cdot 10^{-8}\, M_\odot/\textrm{yr}$ to nicely exemplify all effects and $\alpha$ is set to $0.95 \cdot 10^{-2}$, cases with different $\alpha$ and $\dot{M}$ are explored in the next section. Little is known about the turbulent Prandtl number in accretion disks, but the Prandtl number of hydrogen at temperatures from 700\,K to 1000\,K is 0.68 \cite{H2} and it is this value that is used in the model. Furthermore, like $\alpha$, the Prandtl number is taken to be uniform in the computational domain, and the optical depth in the first cell $\tau_0$ is multiplied with $q_\tau = 0.1$ to prevent the rim from moving outside the computational domain \cite{FL16}.

Figure \ref{fig:VHTemp} depicts the midplane temperature profiles of the four runs. 
A significantly hotter inner rim is obtained in the cases with viscous heating, with temperatures reaching the evaporation temperature. Particularly the optically thick regions deviate in temperature from the runs without viscous heating, because the heat generated in these regions cannot be radiated away efficiently. The two runs without viscous heating, (\texttt{TC0VH0}) and (\texttt{TC1VH0}), have almost identical temperature profiles as expected for non viscous heated cases with the same surface density. The temperature in the shadowed region is elevated through the dissipation as well and regions with fully evaporated dust begin to appear at around 0.5\,AU. 

Due to the diffusive nature of the heat conduction the radial variation of the temperature is less steep in the (\texttt{TC1VH1}) case compared to (\texttt{TC1VH0}). 
The fairly pronounced temperature sink before the rise are still present and the shadowed regions are also slightly cooler. In general the effect of thermal conduction effectively cools the disk and leads to a significant difference in the profile.

\subsection{The influence of the accretion rate}\label{thermCond}

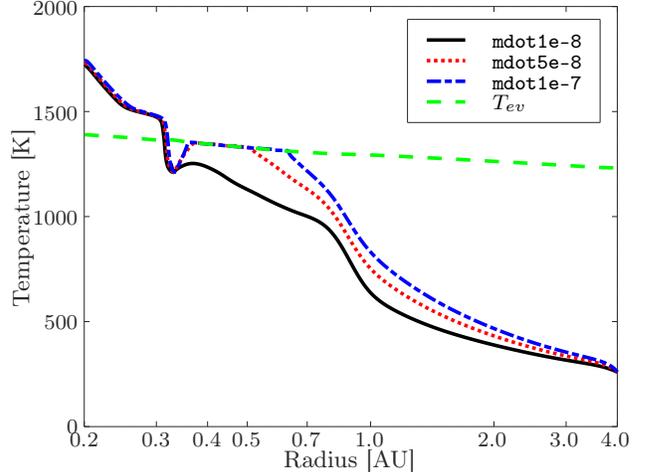
\begin{figure}
	\input{radialtemp3.tex}
	\vspace{-1cm}
	\caption{Comparison of three radial temperature profiles of runs including viscous heating and thermal conduction, but with different accretion rates, the black solid line is for the model \texttt{mdot1e-8}, the red dotted line is for \texttt{mdot5e-8} (note that this is identical to \texttt{TC1VH1}) and the blue dashed and dotted line is for \texttt{mdot1e-7}. The evaporation temperature (green dashed line) is shown for reference.}
	\label{fig:TCTemp}
\end{figure}

\begin{table}
	\caption{\label{tab:radii}Model results for different accretion rates}
	\begin{indented}
		\item[]\begin{tabular}{@{}llllll}
			\br
			Model&$\tau_\mathrm{r}^* = 1$&$T_\textrm{mid}= 1000$K\\
			\mr
			\texttt{mdot1e-8}&0.29&0.70\\
			\texttt{mdot5e-8}&0.29&0.82\\
			\texttt{mdot1e-7}&0.29&0.87\\
			\br
			\multicolumn{3}{l}{\textbf{Note.} Model name, position of dust rim ($\tau_\mathrm{r}^* = 1$) and}\\
			\multicolumn{3}{l}{position of the dead zone edge ($T_\textrm{mid}= 1000$K) in AU}
		\end{tabular}
	\end{indented}
\end{table}

\begin{figure*}
	\input{tempTC.tex}
	\input{dustdensTC.tex}
	\vspace{-0.5cm}
	\caption{2D profiles of temperature (top) and the logarithm of the dust density in g/cm$^3$ to the base 10 (bottom) for the \texttt{mdot1e-7} case at an early stage. The y-axis is the polar angle in rad offset by $\pi/2$ and the x-axis the radial distance in AU.}
	\label{fig:struc2d}
\end{figure*}
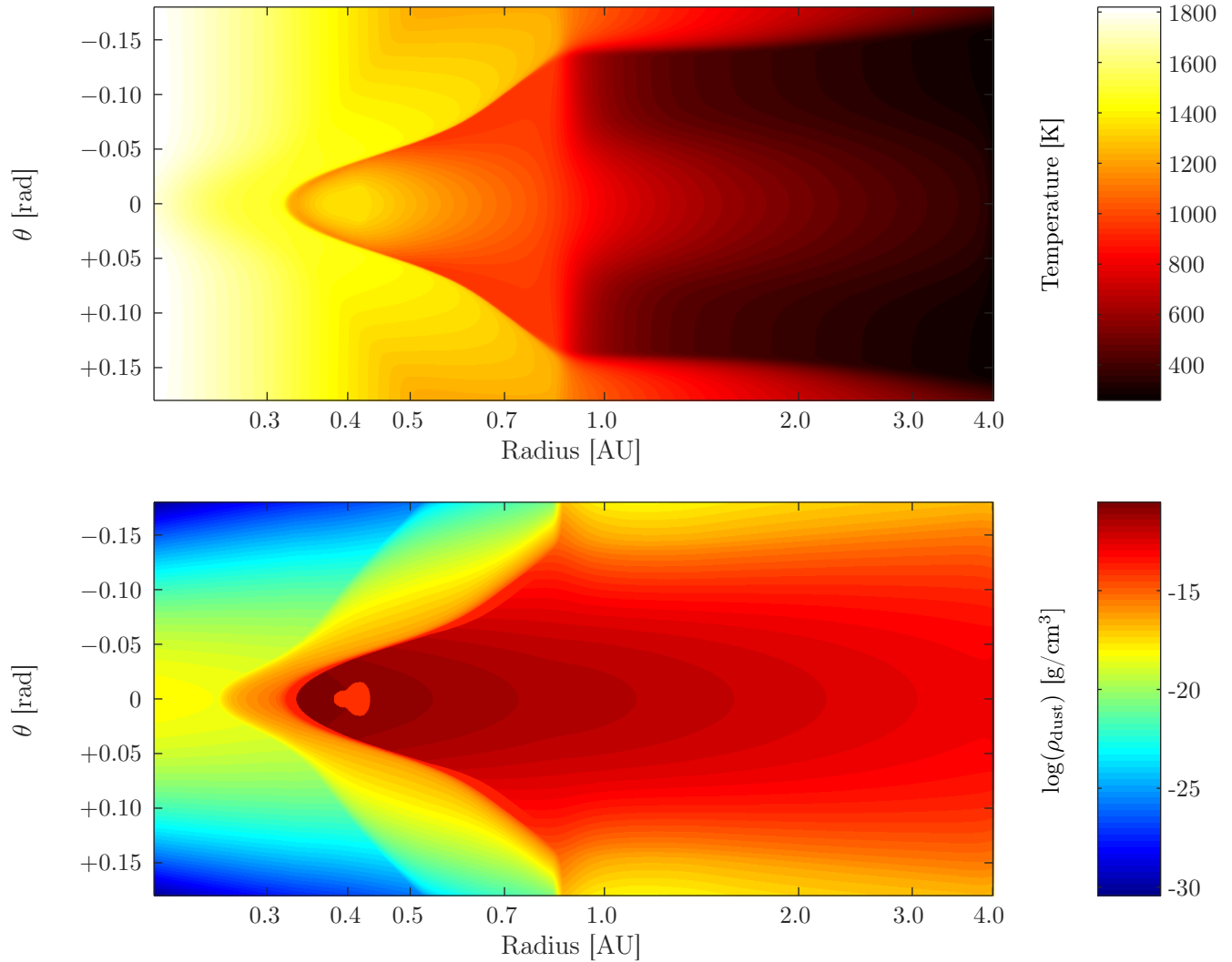

All cases in this section consider both viscous heat as well as thermal conduction. All of them have the same surface density (provided that temperatures are equal), but different accretion rates and corresponding $\alpha$.
Three cases are considered with parameters identical to those listed in table \ref{tab:viscHeat}.

Since typical accretion rates for Herbig Ae/Be stars are expected to be high, especially in the early stages, the effect of viscous heating is important. Experimental data suggest that accretion rates up to $10^{-6} \, M_\odot/\textrm{yr}$ are common \cite{vdA04} and that 25\% of HAe stars have $\dot{M} > 10^{-7} \, M_\odot/\textrm{yr}$ \cite{LNT}. For this reason, $\dot{M} = 10^{-7} \, M_\odot/\textrm{yr}$, $\dot{M} = 5 \cdot 10^{-8} \, M_\odot/\textrm{yr}$ and $\dot{M} = 10^{-8} \, M_\odot/\textrm{yr}$ are chosen for models (\texttt{mdot1e-7}), (\texttt{mdot5e-8}) and (\texttt{mdot1e-8}), respectively.

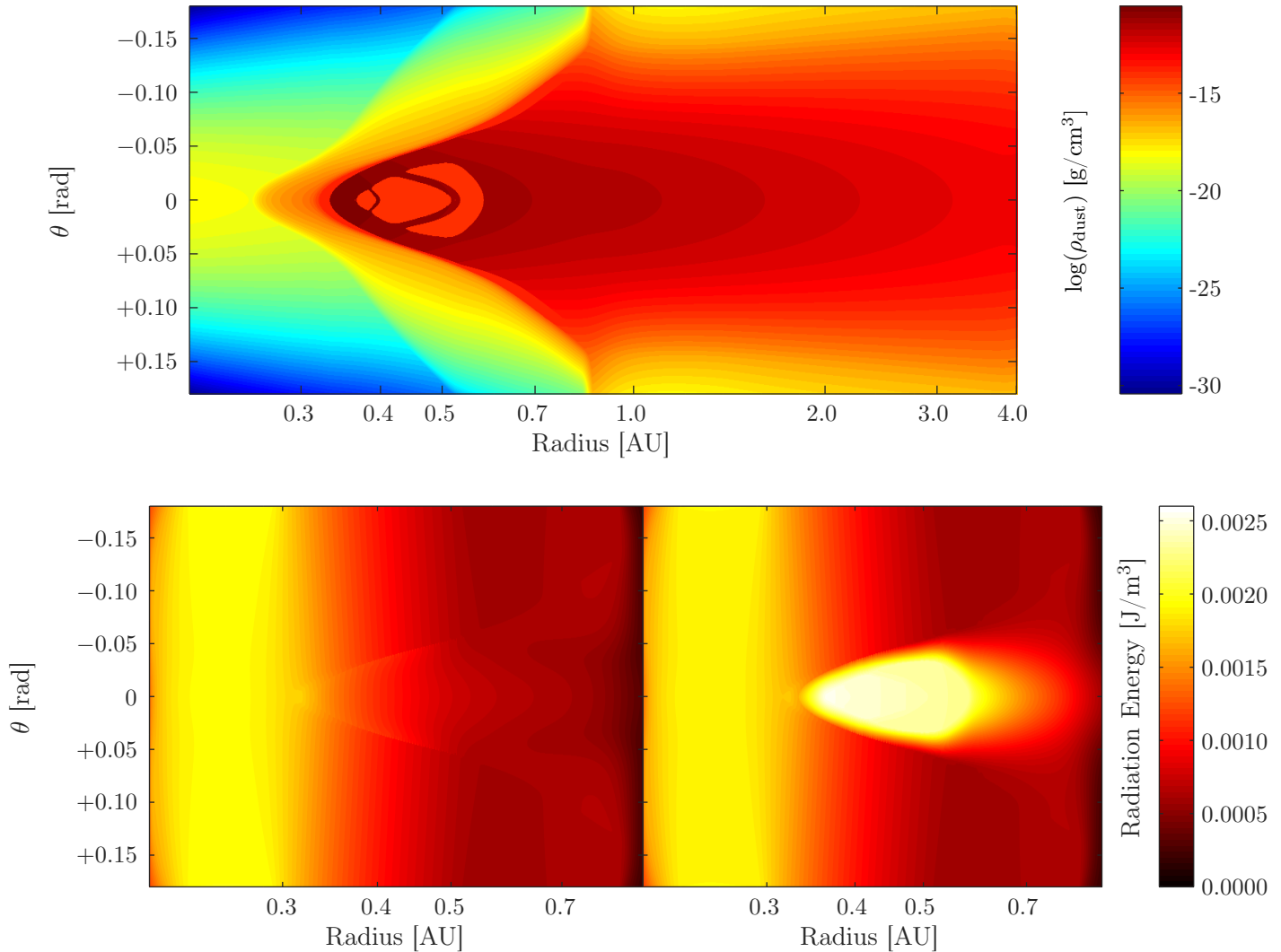
\begin{figure*}
	\input{dustdensTC2.tex}
	\input{Er.tex}
	\vspace{-0.5cm}
	\caption{2D profile of the logarithm of the dust density in g/cm$^3$ to the base 10 (top) for the \texttt{mdot1e-7} case at a late stage with waves visible and a comparison of the radiation energy densities (bottom) of the \texttt{jTC1e-7} case (left) and the \texttt{mdot1e-7} case. The y-axis is the polar angle in rad offset by $\pi/2$ and the x-axis the radial distance in AU.}
	\label{fig:waves}
\end{figure*}

In order for all cases to have the same surface density, $\alpha$ is chosen to be $1.9 \cdot 10^{-2}$, $\alpha = 0.95 \cdot 10^{-2}$ and $1.9 \cdot 10^{-3}$ for the (\texttt{mdot1e-7}), (\texttt{mdot5e-8}) and (\texttt{mdot1e-8}) case, respectively. 
These represent realistic values for thin disks \cite{KPL,DSP}. Furthermore, $\alpha$ is constant throughout the whole box.

The midplane temperature profiles of all three runs are depicted in figure \ref{fig:TCTemp}. 
For the \texttt{mdot1e-8} case, the temperature no longer reaches the evaporation temperature, but the other two still have an area where the dust is fully evaporated. The area, where all the dust is evaporated has a wider extend in the case with an higher accretion rate.

These gaseous regions could be ionized and MRI active, thus vindicating the high values for $\alpha$, that are used in the setup. Although the temperature at which the MRI sets in depends on various parameters, like gas density and dust-to-gas ratio \cite{UN,DT}, a typical value is 1000\,K. Both runs with viscous heating turned on are well above this mark before 1\,AU.

The position of the dust rim ($\tau_\mathrm{r}^* = 1$) does not change for different accretion rates, as seen in table \ref{tab:radii}. For all three cases the rim lies at $0.29$ AU, since its position is determined by the surface density in the gaseous disk and these are identical for all three models.

The dead zone edge ($T_\textrm{mid}= 1000$K), the point where the MRI is no longer dominant lies further from the star as the accretion rate increases. The range is $0.70 - 0.87$ AU and especially the value for \texttt{mdot1e-7} $0.87$ AU is in good agreement with the value that was found in \cite{FL16} for \texttt{MDe-7} of $0.86$ AU. It lies further outward because the viscous heating can maintain a larger hot section for stronger accretion rates.

\subsection{Structure of the rim}\label{sec:struc}

The structure of the rim is affected by the viscous dissipation and turbulent heat conduction. It still displays four distinct regions as found by \cite{FL16} but has a very different thermal and density profile as can be seen in figure \ref{fig:struc2d} for the (\texttt{mdot1e-7}) case. Depicted is the (\texttt{mdot1e-7}) case with parameters as in table \ref{tab:viscHeat} but $N_2 = 1$ and $N_3 = 135$. This is the case at an earlier stage than previously to show the development of the dust free zone inside the rounded off rim.

The first region is dust free inner zone that follows the optically thin gas temperature and lies inward of 0.25\,AU.

The next zone is the dust halo where the dust-to-gas ratio gradually rises before the condensation temperature is reached at 0.35\,AU.

Following that is the condensation front that engulfs the third region. This section is formed by the rounded off rim and in its centre can be an active zone that reaches the condensation temperature through the viscous dissipation. This is the exact opposite of a dead zone, a zone where temperature is too low for MRI to arise. This region can reach MRI capable conditions and is kept at high temperature because of the resulting high viscosity in combination with the high optical thickness that does not allow for effective radiative cooling. 

The last and fourth region is the outer disk that is getting colder with radial distance and lies at optical depth beyond unity. The part of this region that is close to the equatorial plane is warmer than the higher altitudes, the outer disk is not isothermal at constant radius, contrary to the  case without viscous heating.

Potentially there are two stable states that can be reached by an protoplanetary disk: one with a dead zone, if the viscous heating is insufficient and the area inside the rounded off rim cools because no external radiation can heat it and another with and active zone, if the viscous heating is strong enough and the heat is trapped inside the disk. These two cases can be seen in the lower depiction of figure \ref{fig:waves}. On the left viscous heating is disabled and the radiation energy density inside the disk is almost in equilibrium with the radiation energy density outside. On the right viscous heating is included and a significant increase of radiation energy density inside the disk is visible. This means the energy created by the viscous heating cannot be radiated away at a sufficient rate through the optically thick rim. That leads to an equilibrium with an active zone. Both cases use parameters as in table \ref{tab:viscHeat} but $N_2 = 3$ and $N_3 = 170$. 

Interesting to note is that despite the strong radial temperature gradients no local pressure maximum forms in the midplane. In agreement with \cite{FL16} there is no local pressure maximum for constant $\alpha$, once a temperature dependent $\alpha$ is introduced a local pressure maximum appears at the location where the viscosity changes. In the model description of this paper the viscosity is uniform, and a rise in the temperature with radius occurs through the viscous heating. This increase in temperature, however, is insufficient for the pressure gradient to change sign. 

The upper depiction of figure \ref{fig:waves} shows the (\texttt{mdot1e-7}) case with parameters as in table \ref{tab:viscHeat} but $N_2 = 3$ and $N_3 = 170$. It shows that the active zone inside the rounded off rim in its final size. After the active zone has reached a certain size small temperature deviations start to occur and travel through it. These small deviations manifest in the depiction as walls where the dust is present. Since the simulation does not include latent heat it is not possible to decide whether these small deviations have to occur or not, but they were consistently found for every parameter set that produced a dust free zone.

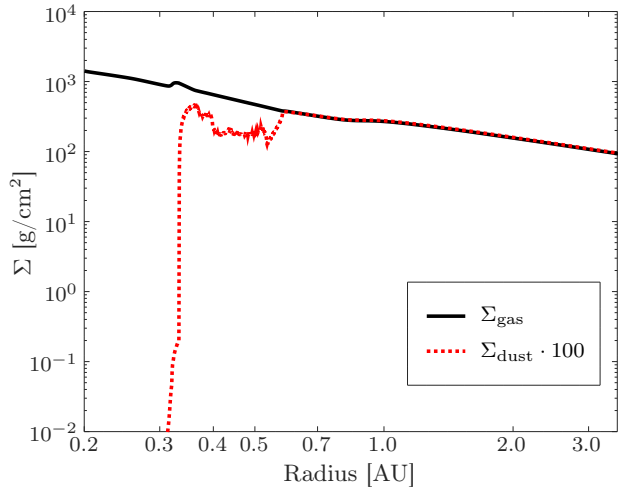
\begin{figure}
	\input{radialdens.tex}
	\vspace{-0.5cm}
	\caption{Comparison of gas (solid line) and dust (dotted line) surface density for model \texttt{mdot1e-7}.}
	\label{fig:radialdens}
\end{figure}
\begin{figure}
	\input{tau.tex}
	\vspace{-0.5cm}
	\caption{Vertical optical depth in NIR for model \texttt{mdot1e-7}.}
	\label{fig:tauz}
\end{figure}
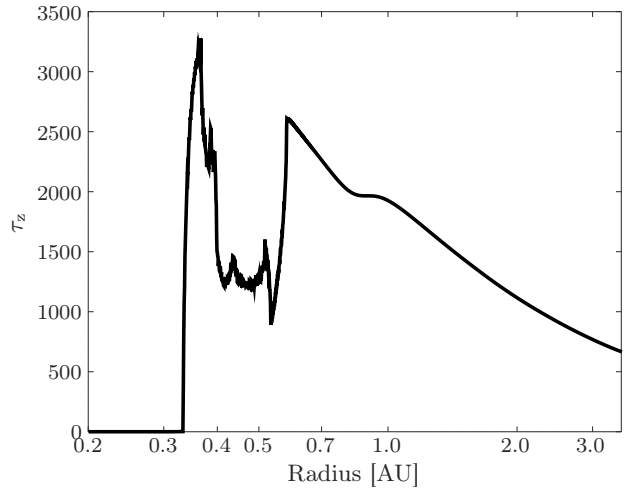

The surface density in figure \ref{fig:radialdens} mirrors the cavity that is also present in \ref{fig:waves}. The dust density falls to $\Sigma_\textrm{dust} \approx 2$ g/cm$^2$ in the dust free zone. To further analyse the equilibrium in the dust free zone an alternative sublimation formula is in progress.
	
The optical depth at constant radius
\begin{equation}
	\tau_\mathrm{z}= \int\limits_{-\infty}^{\infty} \sigma dz
\end{equation} 
for near infrared radiation in figure \ref{fig:tauz} suggests that the dust free zone forms after a large enough depth in $\theta$ direction is reached, because diffusion of radiation energy is no longer sufficient to cool the disk.

\section{Discussion}

This work represents an early stage in developing a self-consistent model of the inner rim of protoplanetary disks. In this section some important limitations of the model will be discussed.

The simplifications introduced in \cite{FL16}, such as a uniform dust-to-gas ratio in the shadowed region and a non-frequency-dependent gas opacity are also applied in this work. A more elaborate sublimation formula is needed in the future, especially with respect to the dust free cavity inside the disk. A detailed implementation of gas line radiation transfer would go beyond the scope of this work. Further the value of $\alpha$ was chosen constant throughout the box. However, the strength of the turbulence is expected be influenced by the temperature and could be lower in MRI inactive regions \cite{Lesur,Tur,Simon}. A temperature dependent $\alpha$ will be explored in further work.

The temperature deviations and dust walls described at the end of section \ref{sec:struc} will have to be further scrutinized in combination with the modelling of the dust-to-gas ratio. Small differences in the temperature can lead to significant changes in the dust-to-gas ratio around the sublimation temperature. This is therefore a delicate problem that needs to be explored tentatively and might necessitate the inclusion of further effects like dust diffusion, latent heat and temperature dependent viscosity.

\section{Conclusions and outlook}

This paper presents a 2D model for protoplanetary disks that consistently includes viscous heating and thermal conduction. The model expands previous works \cite{FL16} and explores new parameter ranges that led to qualitatively new results. These results are:
\begin{enumerate}[label=\arabic*)]
	\item For accretion rates $\dot{M}\ge 3 \cdot 10^{-8} M_\odot/\textrm{yr}$ the viscous dissipation can not be neglected. It affects all regions and can cause a MRI active zone to form behind the condensation front.
	\item The thermal conduction is important since high radial temperature gradients exist through the accretion heating. The cooling through thermal conduction is a significant part in the energy balance of the disk, it can therefore also not be neglected for the parameters used.
	\item Despite the strong radial temperature gradients no local pressure maximum forms in the midplane. 
	Therefore, the model calculations do not provide for a region where the radial drift of grains generated by the friction with the gas is reversed \cite{GuiSan}.
	\item The inner rim position is not affected by viscous heating but determined by the surface density.
	\item The dead zone edge shifts radially outward for higher accretion rates because of a larger zone with viscous heating.
\end{enumerate}
The code used to produce these results can be found at \mbox{bitbucket.org/astro\_bayreuth/radiation\_code}.

This model could be expanded through temperature depended viscosity or thermal conductivity and embedded it in a computationally more demanding 3D simulation. Also a subsequent paper is devoted to study the spectral energy densities produced by active disks. But this lies without the scope of this work.

\section*{References}
\bibliography{references.bib}
\end{document}

%% file: radialtemp.tex
\setlength{\unitlength}{0.6pt}
\begin{picture}(0,0)
\includegraphics[scale=0.6]{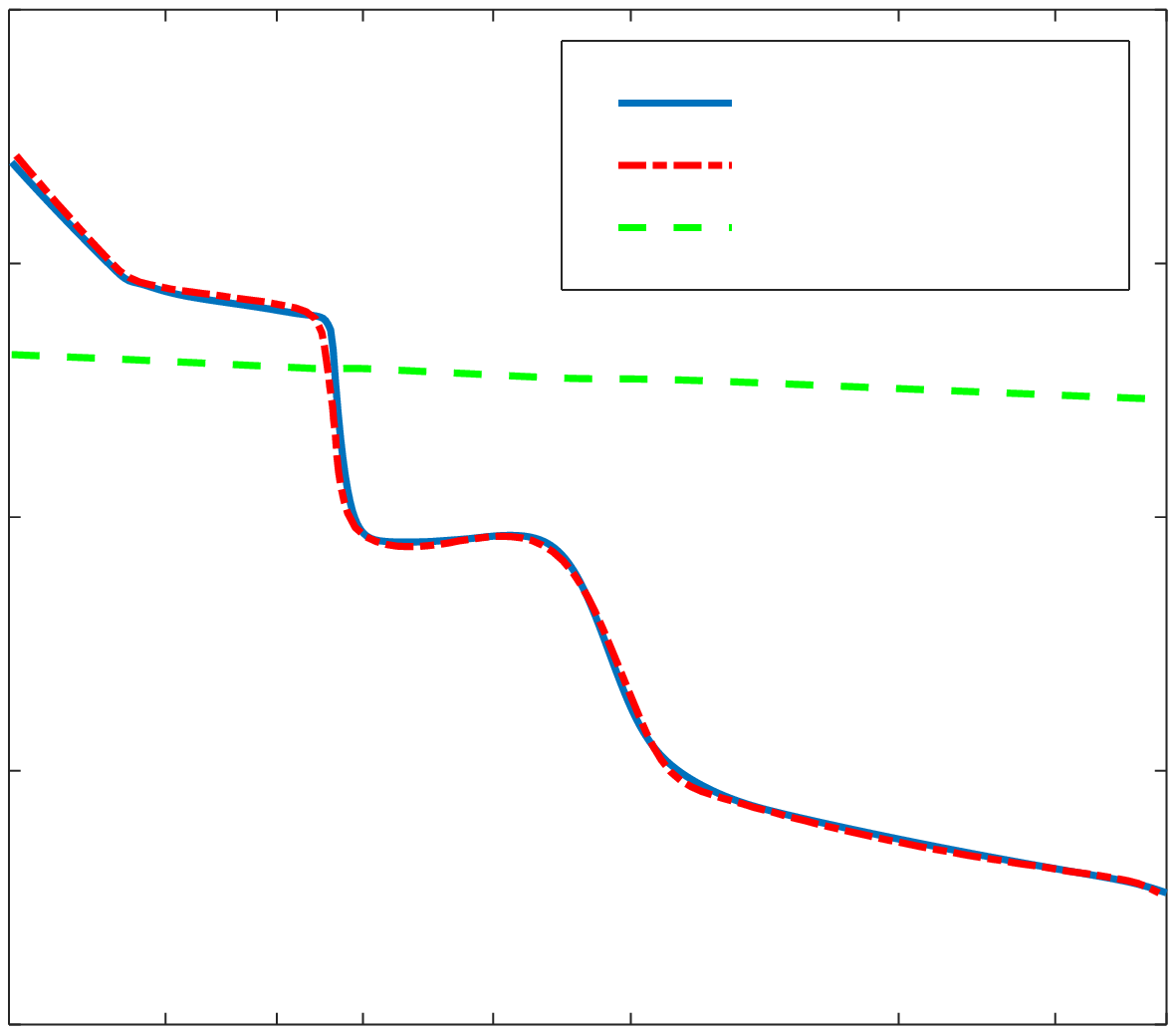}\label{figure1}
\end{picture}%
\begin{picture}(432,360)(0,0)
\fontsize{8}{0}
\selectfont\put(56.1602,34.5933){\makebox(0,0)[t]{\textcolor[rgb]{0.15,0.15,0.15}{{0.2}}}}
\fontsize{8}{0}
\selectfont\put(101.474,34.5933){\makebox(0,0)[t]{\textcolor[rgb]{0.15,0.15,0.15}{{0.3}}}}
\fontsize{8}{0}
\selectfont\put(133.625,34.5933){\makebox(0,0)[t]{\textcolor[rgb]{0.15,0.15,0.15}{{0.4}}}}
\fontsize{8}{0}
\selectfont\put(158.563,34.5933){\makebox(0,0)[t]{\textcolor[rgb]{0.15,0.15,0.15}{{0.5}}}}
\fontsize{8}{0}
\selectfont\put(196.167,34.5933){\makebox(0,0)[t]{\textcolor[rgb]{0.15,0.15,0.15}{{0.7}}}}
\fontsize{8}{0}
\selectfont\put(236.029,34.5933){\makebox(0,0)[t]{\textcolor[rgb]{0.15,0.15,0.15}{{1.0}}}}
\fontsize{8}{0}
\selectfont\put(313.495,34.5933){\makebox(0,0)[t]{\textcolor[rgb]{0.15,0.15,0.15}{{2.0}}}}
\fontsize{8}{0}
\selectfont\put(358.809,34.5933){\makebox(0,0)[t]{\textcolor[rgb]{0.15,0.15,0.15}{{3.0}}}}
\fontsize{8}{0}
\selectfont\put(390.96,34.5933){\makebox(0,0)[t]{\textcolor[rgb]{0.15,0.15,0.15}{{4.0}}}}
\fontsize{8}{0}
\selectfont\put(51.1631,39.6001){\makebox(0,0)[r]{\textcolor[rgb]{0.15,0.15,0.15}{{0}}}}
\fontsize{8}{0}
\selectfont\put(51.1631,112.95){\makebox(0,0)[r]{\textcolor[rgb]{0.15,0.15,0.15}{{500}}}}
\fontsize{8}{0}
\selectfont\put(51.1631,186.3){\makebox(0,0)[r]{\textcolor[rgb]{0.15,0.15,0.15}{{1000}}}}
\fontsize{8}{0}
\selectfont\put(51.1631,259.65){\makebox(0,0)[r]{\textcolor[rgb]{0.15,0.15,0.15}{{1500}}}}
\fontsize{8}{0}
\selectfont\put(51.1631,333){\makebox(0,0)[r]{\textcolor[rgb]{0.15,0.15,0.15}{{2000}}}}
\fontsize{9}{0}
\selectfont\put(223.56,23.5933){\makebox(0,0)[t]{\textcolor[rgb]{0.15,0.15,0.15}{{Radius [AU]}}}}
\fontsize{9}{0}
\selectfont\put(23.1631,186.3){\rotatebox{90}{\makebox(0,0)[b]{\textcolor[rgb]{0.15,0.15,0.15}{{Temperature [K]}}}}}
\fontsize{8}{0}
\selectfont\put(289.872,306){\makebox(0,0)[l]{\textcolor[rgb]{0,0,0}{{Schobert et al.}}}}
\fontsize{8}{0}
\selectfont\put(289.872,288){\makebox(0,0)[l]{\textcolor[rgb]{0,0,0}{{Flock et al.}}}}
\fontsize{8}{0}
\selectfont\put(289.872,270){\makebox(0,0)[l]{\textcolor[rgb]{0,0,0}{{$T_{\mathrm{ev}}$}}}}
\end{picture}

%% file: temp.tex
\setlength{\unitlength}{1pt}
\begin{picture}(0,0)
\includegraphics{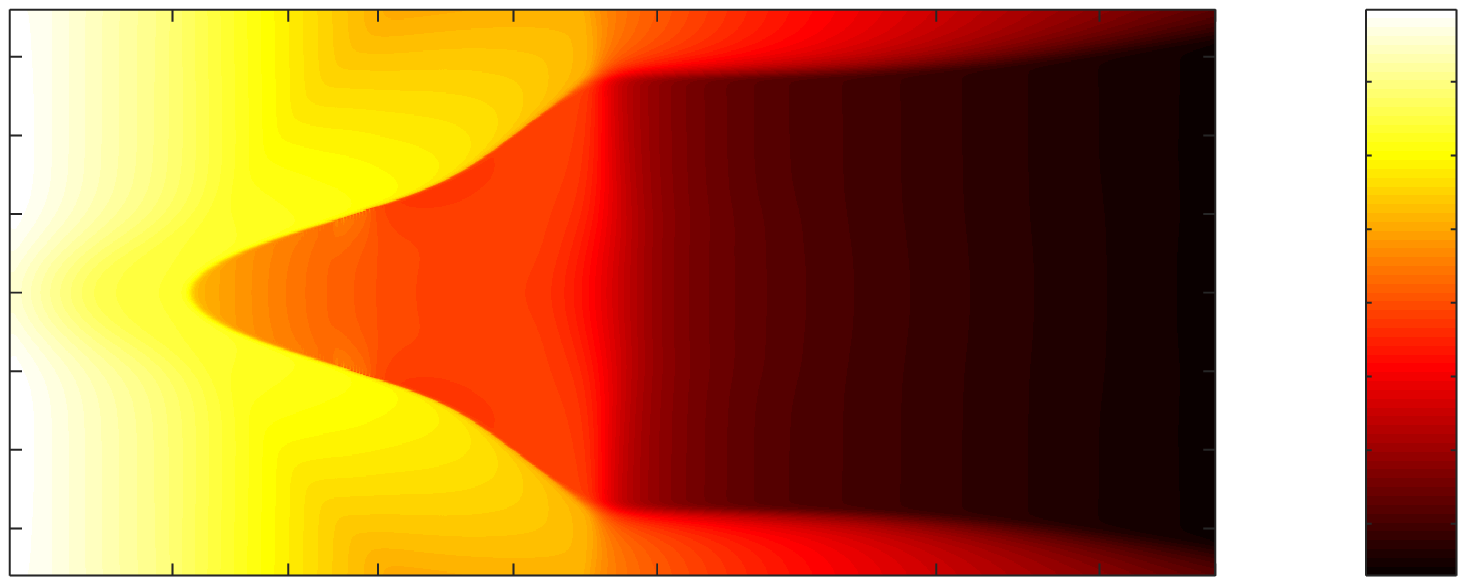}
\end{picture}%
\begin{picture}(560,200)(0,0)
\fontsize{10}{0}
\selectfont\put(446.092,103.5){\rotatebox{90}{\makebox(0,0){\textcolor[rgb]{0,0,0}{{Temperature [K]}}}}}
\fontsize{10}{0}
\selectfont\put(119.155,17){\makebox(0,0)[t]{\textcolor[rgb]{0.15,0.15,0.15}{{0.3}}}}
\fontsize{10}{0}
\selectfont\put(152.558,17){\makebox(0,0)[t]{\textcolor[rgb]{0.15,0.15,0.15}{{0.4}}}}
\fontsize{10}{0}
\selectfont\put(178.466,17){\makebox(0,0)[t]{\textcolor[rgb]{0.15,0.15,0.15}{{0.5}}}}
\fontsize{10}{0}
\selectfont\put(217.534,17){\makebox(0,0)[t]{\textcolor[rgb]{0.15,0.15,0.15}{{0.7}}}}
\fontsize{10}{0}
\selectfont\put(258.947,17){\makebox(0,0)[t]{\textcolor[rgb]{0.15,0.15,0.15}{{1.0}}}}
\fontsize{10}{0}
\selectfont\put(339.427,17){\makebox(0,0)[t]{\textcolor[rgb]{0.15,0.15,0.15}{{2.0}}}}
\fontsize{10}{0}
\selectfont\put(386.505,17){\makebox(0,0)[t]{\textcolor[rgb]{0.15,0.15,0.15}{{3.0}}}}
\fontsize{10}{0}
\selectfont\put(419.908,17){\makebox(0,0)[t]{\textcolor[rgb]{0.15,0.15,0.15}{{4.0}}}}
\fontsize{10}{0}
\selectfont\put(67.7969,171.417){\makebox(0,0)[r]{\textcolor[rgb]{0.15,0.15,0.15}{{$-0.15$}}}}
\fontsize{10}{0}
\selectfont\put(67.7969,148.778){\makebox(0,0)[r]{\textcolor[rgb]{0.15,0.15,0.15}{{$-0.10$}}}}
\fontsize{10}{0}
\selectfont\put(67.7969,126.139){\makebox(0,0)[r]{\textcolor[rgb]{0.15,0.15,0.15}{{$-0.05$}}}}
\fontsize{10}{0}
\selectfont\put(67.7969,103.5){\makebox(0,0)[r]{\textcolor[rgb]{0.15,0.15,0.15}{{$0$}}}}
\fontsize{10}{0}
\selectfont\put(67.7969,80.8608){\makebox(0,0)[r]{\textcolor[rgb]{0.15,0.15,0.15}{{$+0.05$}}}}
\fontsize{10}{0}
\selectfont\put(67.7969,58.2222){\makebox(0,0)[r]{\textcolor[rgb]{0.15,0.15,0.15}{{$+0.10$}}}}
\fontsize{10}{0}
\selectfont\put(67.7969,35.5835){\makebox(0,0)[r]{\textcolor[rgb]{0.15,0.15,0.15}{{$+0.15$}}}}
\fontsize{11}{0}
\selectfont\put(246.4,6){\makebox(0,0)[t]{\textcolor[rgb]{0.15,0.15,0.15}{{Radius [AU]}}}}
\fontsize{11}{0}
\selectfont\put(24.7969,103.5){\rotatebox{90}{\makebox(0,0)[b]{\textcolor[rgb]{0.15,0.15,0.15}{{$\theta$ [rad]}}}}}
\fontsize{10}{0}
\selectfont\put(494.448,36.9058){\makebox(0,0)[l]{\textcolor[rgb]{0.15,0.15,0.15}{{400}}}}
\fontsize{10}{0}
\selectfont\put(494.448,58.1235){\makebox(0,0)[l]{\textcolor[rgb]{0.15,0.15,0.15}{{600}}}}
\fontsize{10}{0}
\selectfont\put(494.448,79.3418){\makebox(0,0)[l]{\textcolor[rgb]{0.15,0.15,0.15}{{800}}}}
\fontsize{10}{0}
\selectfont\put(494.448,100.56){\makebox(0,0)[l]{\textcolor[rgb]{0.15,0.15,0.15}{{1000}}}}
\fontsize{10}{0}
\selectfont\put(494.448,121.777){\makebox(0,0)[l]{\textcolor[rgb]{0.15,0.15,0.15}{{1200}}}}
\fontsize{10}{0}
\selectfont\put(494.448,142.996){\makebox(0,0)[l]{\textcolor[rgb]{0.15,0.15,0.15}{{1400}}}}
\fontsize{10}{0}
\selectfont\put(494.448,164.213){\makebox(0,0)[l]{\textcolor[rgb]{0.15,0.15,0.15}{{1600}}}}
\end{picture}

%% file: dustdens.tex
\setlength{\unitlength}{1pt}
\begin{picture}(0,0)
\includegraphics{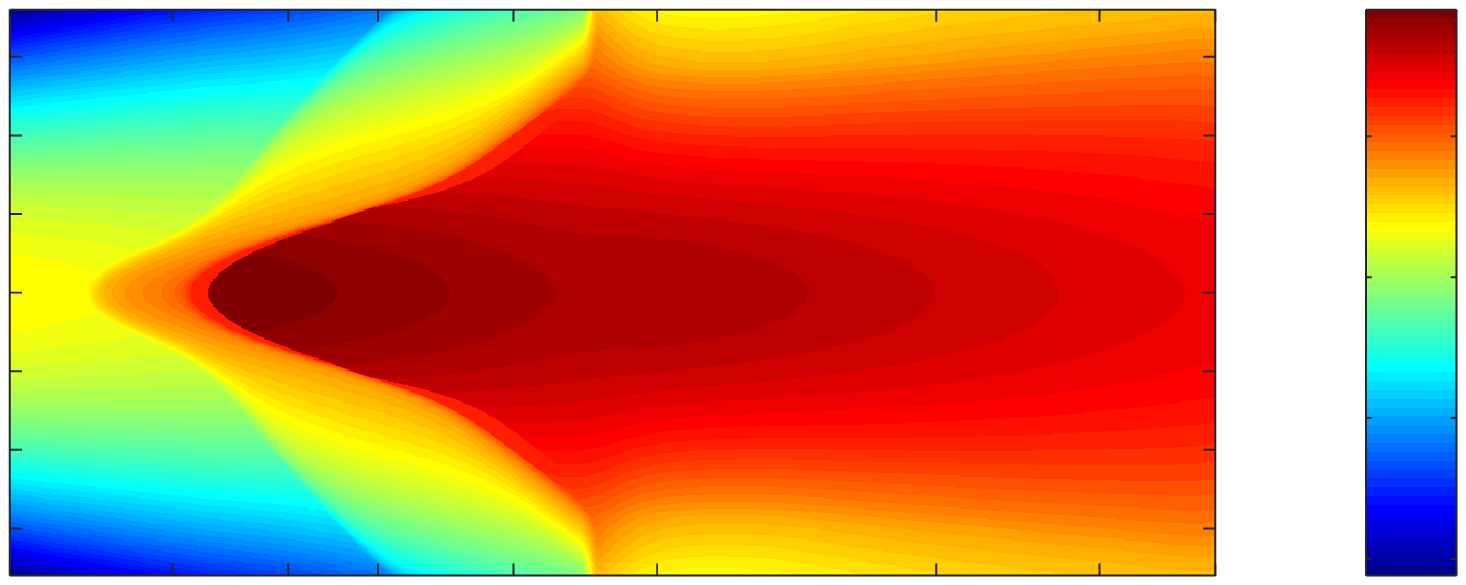}
\end{picture}%
\begin{picture}(560,205)(0,0)
\fontsize{10}{0}
\selectfont\put(446.092,103.5){\rotatebox{90}{\makebox(0,0){\textcolor[rgb]{0,0,0}{{log($\rho_{\mathrm{dust}}$) [g/cm$^3]$}}}}}
\fontsize{10}{0}
\selectfont\put(119.155,17){\makebox(0,0)[t]{\textcolor[rgb]{0.15,0.15,0.15}{{0.3}}}}
\fontsize{10}{0}
\selectfont\put(152.558,17){\makebox(0,0)[t]{\textcolor[rgb]{0.15,0.15,0.15}{{0.4}}}}
\fontsize{10}{0}
\selectfont\put(178.466,17){\makebox(0,0)[t]{\textcolor[rgb]{0.15,0.15,0.15}{{0.5}}}}
\fontsize{10}{0}
\selectfont\put(217.534,17){\makebox(0,0)[t]{\textcolor[rgb]{0.15,0.15,0.15}{{0.7}}}}
\fontsize{10}{0}
\selectfont\put(258.947,17){\makebox(0,0)[t]{\textcolor[rgb]{0.15,0.15,0.15}{{1.0}}}}
\fontsize{10}{0}
\selectfont\put(339.427,17){\makebox(0,0)[t]{\textcolor[rgb]{0.15,0.15,0.15}{{2.0}}}}
\fontsize{10}{0}
\selectfont\put(386.505,17){\makebox(0,0)[t]{\textcolor[rgb]{0.15,0.15,0.15}{{3.0}}}}
\fontsize{10}{0}
\selectfont\put(419.908,17){\makebox(0,0)[t]{\textcolor[rgb]{0.15,0.15,0.15}{{4.0}}}}
\fontsize{10}{0}
\selectfont\put(67.7969,171.417){\makebox(0,0)[r]{\textcolor[rgb]{0.15,0.15,0.15}{{$-0.15$}}}}
\fontsize{10}{0}
\selectfont\put(67.7969,148.778){\makebox(0,0)[r]{\textcolor[rgb]{0.15,0.15,0.15}{{$-0.10$}}}}
\fontsize{10}{0}
\selectfont\put(67.7969,126.139){\makebox(0,0)[r]{\textcolor[rgb]{0.15,0.15,0.15}{{$-0.05$}}}}
\fontsize{10}{0}
\selectfont\put(67.7969,103.5){\makebox(0,0)[r]{\textcolor[rgb]{0.15,0.15,0.15}{{$0$}}}}
\fontsize{10}{0}
\selectfont\put(67.7969,80.8608){\makebox(0,0)[r]{\textcolor[rgb]{0.15,0.15,0.15}{{$+0.05$}}}}
\fontsize{10}{0}
\selectfont\put(67.7969,58.2222){\makebox(0,0)[r]{\textcolor[rgb]{0.15,0.15,0.15}{{$+0.10$}}}}
\fontsize{10}{0}
\selectfont\put(67.7969,35.5835){\makebox(0,0)[r]{\textcolor[rgb]{0.15,0.15,0.15}{{$+0.15$}}}}
\fontsize{11}{0}
\selectfont\put(246.4,6){\makebox(0,0)[t]{\textcolor[rgb]{0.15,0.15,0.15}{{Radius [AU]}}}}
\fontsize{11}{0}
\selectfont\put(24.7969,103.5){\rotatebox{90}{\makebox(0,0)[b]{\textcolor[rgb]{0.15,0.15,0.15}{{$\theta$ [rad]}}}}}
\fontsize{10}{0}
\selectfont\put(494.448,26.7749){\makebox(0,0)[l]{\textcolor[rgb]{0.15,0.15,0.15}{{-30}}}}
\fontsize{10}{0}
\selectfont\put(494.448,67.3643){\makebox(0,0)[l]{\textcolor[rgb]{0.15,0.15,0.15}{{-25}}}}
\fontsize{10}{0}
\selectfont\put(494.448,107.954){\makebox(0,0)[l]{\textcolor[rgb]{0.15,0.15,0.15}{{-20}}}}
\fontsize{10}{0}
\selectfont\put(494.448,148.543){\makebox(0,0)[l]{\textcolor[rgb]{0.15,0.15,0.15}{{-15}}}}
\end{picture}

%% file: radialtemp2.tex
\setlength{\unitlength}{.45pt}
\begin{picture}(0,0)
\includegraphics[scale=0.45]{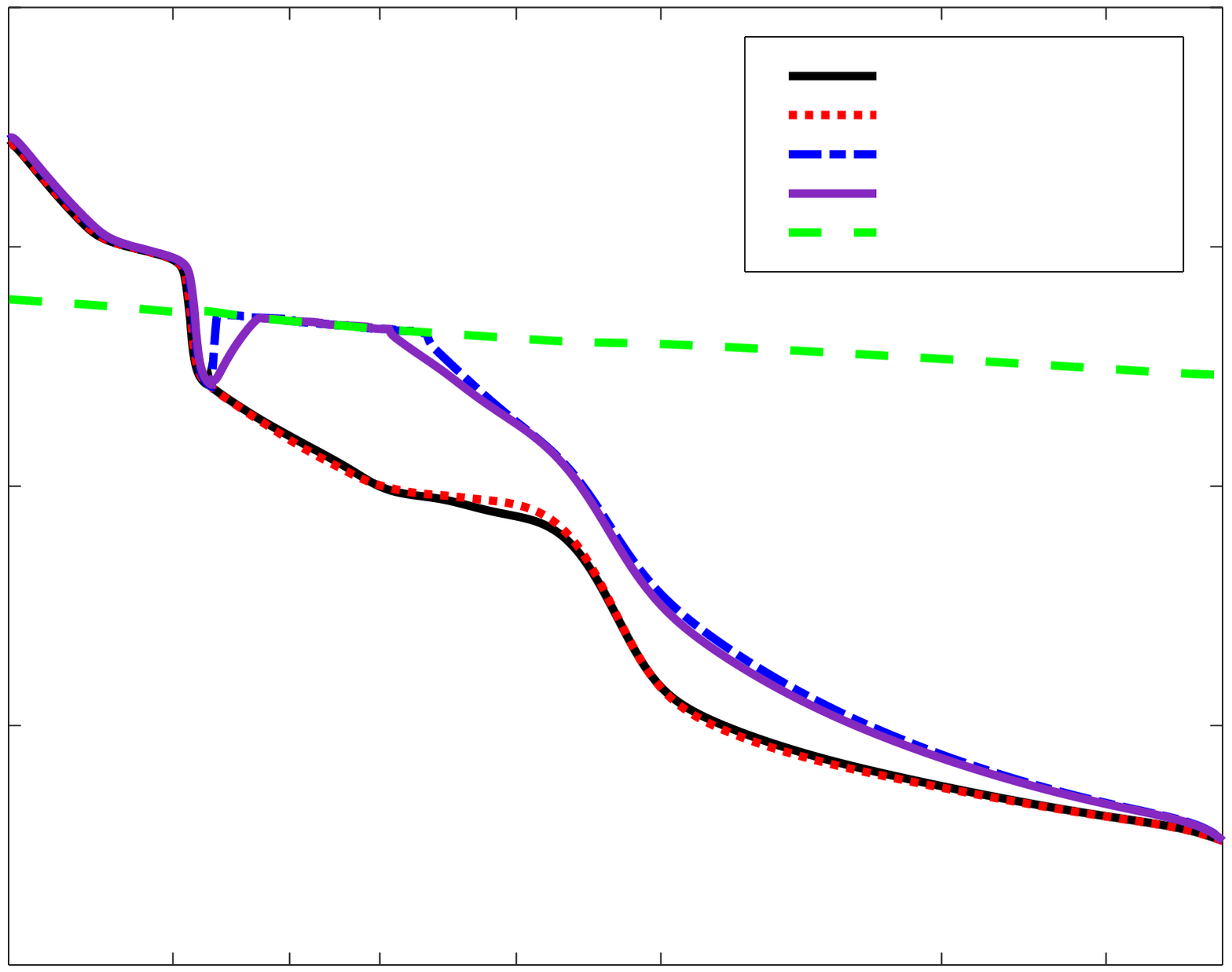}
\end{picture}%
\begin{picture}(576,432)(0,0)
\fontsize{8}{0}
\selectfont\put(74.8799,42.519){\makebox(0,0)[t]{\textcolor[rgb]{0.15,0.15,0.15}{{0.2}}}}
\fontsize{8}{0}
\selectfont\put(135.299,42.519){\makebox(0,0)[t]{\textcolor[rgb]{0.15,0.15,0.15}{{0.3}}}}
\fontsize{8}{0}
\selectfont\put(178.167,42.519){\makebox(0,0)[t]{\textcolor[rgb]{0.15,0.15,0.15}{{0.4}}}}
\fontsize{8}{0}
\selectfont\put(211.418,42.519){\makebox(0,0)[t]{\textcolor[rgb]{0.15,0.15,0.15}{{0.5}}}}
\fontsize{8}{0}
\selectfont\put(261.557,42.519){\makebox(0,0)[t]{\textcolor[rgb]{0.15,0.15,0.15}{{0.7}}}}
\fontsize{8}{0}
\selectfont\put(314.706,42.519){\makebox(0,0)[t]{\textcolor[rgb]{0.15,0.15,0.15}{{1.0}}}}
\fontsize{8}{0}
\selectfont\put(417.993,42.519){\makebox(0,0)[t]{\textcolor[rgb]{0.15,0.15,0.15}{{2.0}}}}
\fontsize{8}{0}
\selectfont\put(478.412,42.519){\makebox(0,0)[t]{\textcolor[rgb]{0.15,0.15,0.15}{{3.0}}}}
\fontsize{8}{0}
\selectfont\put(521.28,42.519){\makebox(0,0)[t]{\textcolor[rgb]{0.15,0.15,0.15}{{4.0}}}}
\fontsize{8}{0}
\selectfont\put(69.8755,47.52){\makebox(0,0)[r]{\textcolor[rgb]{0.15,0.15,0.15}{{0}}}}
\fontsize{8}{0}
\selectfont\put(69.8755,135.54){\makebox(0,0)[r]{\textcolor[rgb]{0.15,0.15,0.15}{{500}}}}
\fontsize{8}{0}
\selectfont\put(69.8755,223.56){\makebox(0,0)[r]{\textcolor[rgb]{0.15,0.15,0.15}{{1000}}}}
\fontsize{8}{0}
\selectfont\put(69.8755,311.58){\makebox(0,0)[r]{\textcolor[rgb]{0.15,0.15,0.15}{{1500}}}}
\fontsize{8}{0}
\selectfont\put(69.8755,399.6){\makebox(0,0)[r]{\textcolor[rgb]{0.15,0.15,0.15}{{2000}}}}
\fontsize{9}{0}
\selectfont\put(298.08,28.519){\makebox(0,0)[t]{\textcolor[rgb]{0.15,0.15,0.15}{{Radius [AU]}}}}
\fontsize{9}{0}
\selectfont\put(33.8755,223.56){\rotatebox{90}{\makebox(0,0)[b]{\textcolor[rgb]{0.15,0.15,0.15}{{Temperature [K]}}}}}
\fontsize{8}{0}
\selectfont\put(418.176,374.371){\makebox(0,0)[l]{\textcolor[rgb]{0,0,0}{{\texttt{TC0VH0}}}}}
\fontsize{8}{0}
\selectfont\put(418.176,360.029){\makebox(0,0)[l]{\textcolor[rgb]{0,0,0}{{\texttt{TC1VH0}}}}}
\fontsize{8}{0}
\selectfont\put(418.176,345.6){\makebox(0,0)[l]{\textcolor[rgb]{0,0,0}{{\texttt{TC0VH1}}}}}
\fontsize{8}{0}
\selectfont\put(418.176,331.171){\makebox(0,0)[l]{\textcolor[rgb]{0,0,0}{{\texttt{TC1VH1}}}}}
\fontsize{8}{0}
\selectfont\put(418.176,316.829){\makebox(0,0)[l]{\textcolor[rgb]{0,0,0}{{$T_{ev}$}}}}
\end{picture}

%% file: radialtemp3.tex
\setlength{\unitlength}{0.45pt}
\begin{picture}(0,0)
\includegraphics[scale=0.45]{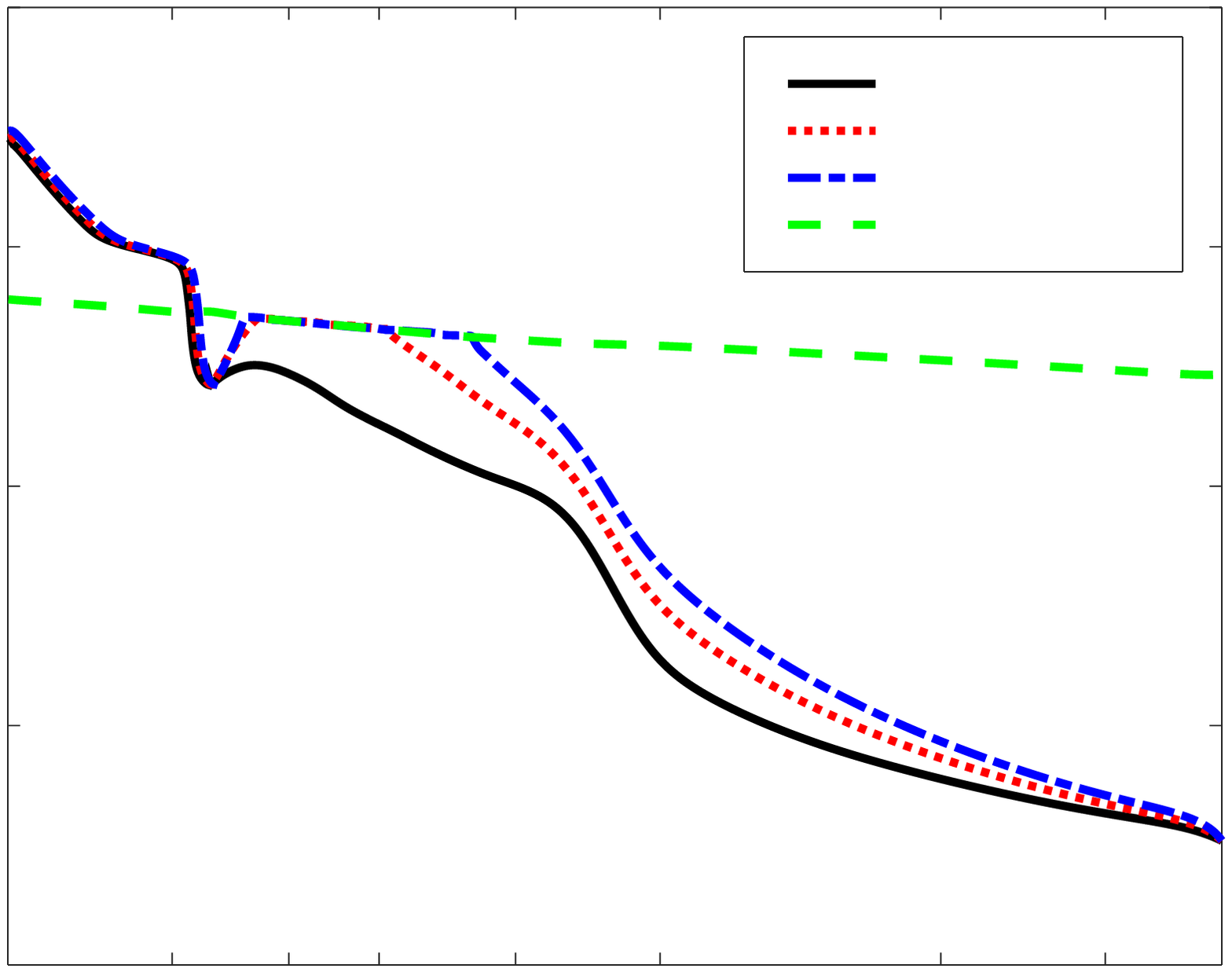}
\end{picture}%
\begin{picture}(576,432)(0,0)
\fontsize{8}{0}
\selectfont\put(74.8799,42.519){\makebox(0,0)[t]{\textcolor[rgb]{0.15,0.15,0.15}{{0.2}}}}
\fontsize{8}{0}
\selectfont\put(135.299,42.519){\makebox(0,0)[t]{\textcolor[rgb]{0.15,0.15,0.15}{{0.3}}}}
\fontsize{8}{0}
\selectfont\put(178.167,42.519){\makebox(0,0)[t]{\textcolor[rgb]{0.15,0.15,0.15}{{0.4}}}}
\fontsize{8}{0}
\selectfont\put(211.418,42.519){\makebox(0,0)[t]{\textcolor[rgb]{0.15,0.15,0.15}{{0.5}}}}
\fontsize{8}{0}
\selectfont\put(261.557,42.519){\makebox(0,0)[t]{\textcolor[rgb]{0.15,0.15,0.15}{{0.7}}}}
\fontsize{8}{0}
\selectfont\put(314.706,42.519){\makebox(0,0)[t]{\textcolor[rgb]{0.15,0.15,0.15}{{1.0}}}}
\fontsize{8}{0}
\selectfont\put(417.993,42.519){\makebox(0,0)[t]{\textcolor[rgb]{0.15,0.15,0.15}{{2.0}}}}
\fontsize{8}{0}
\selectfont\put(478.412,42.519){\makebox(0,0)[t]{\textcolor[rgb]{0.15,0.15,0.15}{{3.0}}}}
\fontsize{8}{0}
\selectfont\put(521.28,42.519){\makebox(0,0)[t]{\textcolor[rgb]{0.15,0.15,0.15}{{4.0}}}}
\fontsize{8}{0}
\selectfont\put(69.8755,47.52){\makebox(0,0)[r]{\textcolor[rgb]{0.15,0.15,0.15}{{0}}}}
\fontsize{8}{0}
\selectfont\put(69.8755,135.54){\makebox(0,0)[r]{\textcolor[rgb]{0.15,0.15,0.15}{{500}}}}
\fontsize{8}{0}
\selectfont\put(69.8755,223.56){\makebox(0,0)[r]{\textcolor[rgb]{0.15,0.15,0.15}{{1000}}}}
\fontsize{8}{0}
\selectfont\put(69.8755,311.58){\makebox(0,0)[r]{\textcolor[rgb]{0.15,0.15,0.15}{{1500}}}}
\fontsize{8}{0}
\selectfont\put(69.8755,399.6){\makebox(0,0)[r]{\textcolor[rgb]{0.15,0.15,0.15}{{2000}}}}
\fontsize{9}{0}
\selectfont\put(298.08,28.519){\makebox(0,0)[t]{\textcolor[rgb]{0.15,0.15,0.15}{{Radius [AU]}}}}
\fontsize{9}{0}
\selectfont\put(33.8755,223.56){\rotatebox{90}{\makebox(0,0)[b]{\textcolor[rgb]{0.15,0.15,0.15}{{Temperature [K]}}}}}
\fontsize{8}{0}
\selectfont\put(418.176,371.52){\makebox(0,0)[l]{\textcolor[rgb]{0,0,0}{{\texttt{mdot1e-8}}}}}
\fontsize{8}{0}
\selectfont\put(418.176,354.24){\makebox(0,0)[l]{\textcolor[rgb]{0,0,0}{{\texttt{mdot5e-8}}}}}
\fontsize{8}{0}
\selectfont\put(418.176,336.96){\makebox(0,0)[l]{\textcolor[rgb]{0,0,0}{{\texttt{mdot1e-7}}}}}
\fontsize{8}{0}
\selectfont\put(418.176,319.68){\makebox(0,0)[l]{\textcolor[rgb]{0,0,0}{{$T_{ev}$}}}}
\end{picture}

%% file: tempTC.tex
\setlength{\unitlength}{1pt}
\begin{picture}(0,0)
\includegraphics{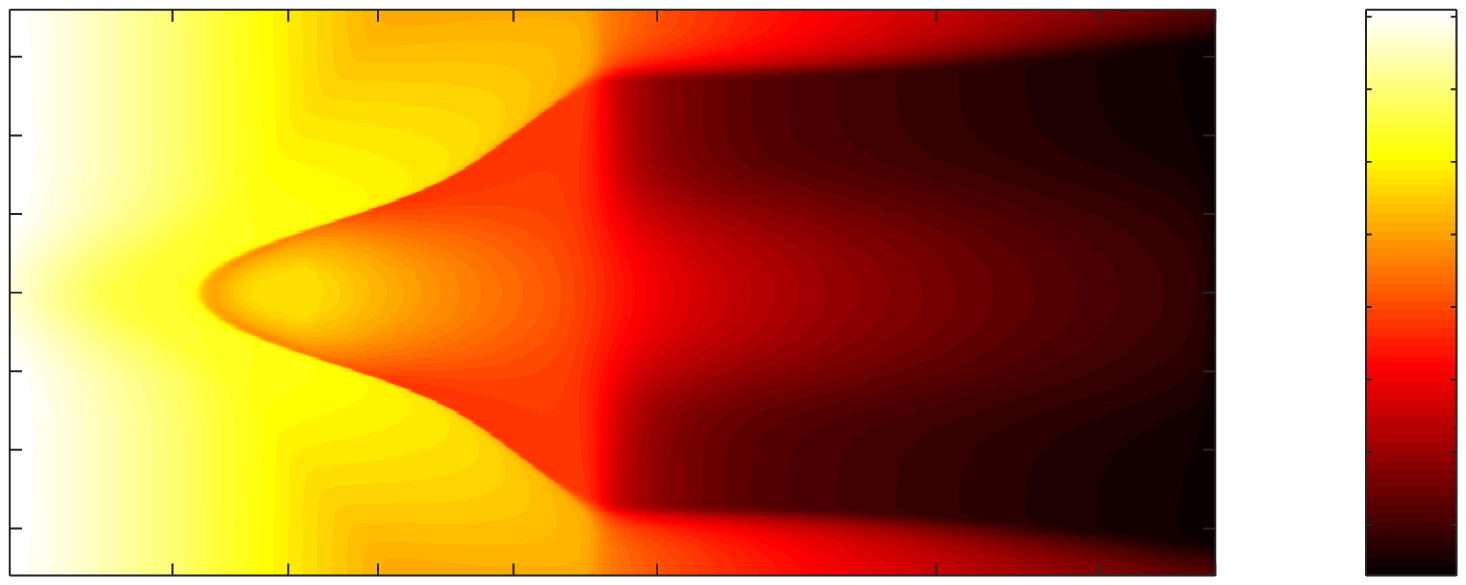}
\end{picture}%
\begin{picture}(560,200)(0,0)
\fontsize{10}{0}
\selectfont\put(446.092,103.5){\rotatebox{90}{\makebox(0,0){\textcolor[rgb]{0,0,0}{{Temperature [K]}}}}}
\fontsize{10}{0}
\selectfont\put(119.155,17){\makebox(0,0)[t]{\textcolor[rgb]{0.15,0.15,0.15}{{0.3}}}}
\fontsize{10}{0}
\selectfont\put(152.558,17){\makebox(0,0)[t]{\textcolor[rgb]{0.15,0.15,0.15}{{0.4}}}}
\fontsize{10}{0}
\selectfont\put(178.466,17){\makebox(0,0)[t]{\textcolor[rgb]{0.15,0.15,0.15}{{0.5}}}}
\fontsize{10}{0}
\selectfont\put(217.534,17){\makebox(0,0)[t]{\textcolor[rgb]{0.15,0.15,0.15}{{0.7}}}}
\fontsize{10}{0}
\selectfont\put(258.947,17){\makebox(0,0)[t]{\textcolor[rgb]{0.15,0.15,0.15}{{1.0}}}}
\fontsize{10}{0}
\selectfont\put(339.427,17){\makebox(0,0)[t]{\textcolor[rgb]{0.15,0.15,0.15}{{2.0}}}}
\fontsize{10}{0}
\selectfont\put(386.505,17){\makebox(0,0)[t]{\textcolor[rgb]{0.15,0.15,0.15}{{3.0}}}}
\fontsize{10}{0}
\selectfont\put(419.908,17){\makebox(0,0)[t]{\textcolor[rgb]{0.15,0.15,0.15}{{4.0}}}}
\fontsize{10}{0}
\selectfont\put(67.7969,171.417){\makebox(0,0)[r]{\textcolor[rgb]{0.15,0.15,0.15}{{$-0.15$}}}}
\fontsize{10}{0}
\selectfont\put(67.7969,148.778){\makebox(0,0)[r]{\textcolor[rgb]{0.15,0.15,0.15}{{$-0.10$}}}}
\fontsize{10}{0}
\selectfont\put(67.7969,126.139){\makebox(0,0)[r]{\textcolor[rgb]{0.15,0.15,0.15}{{$-0.05$}}}}
\fontsize{10}{0}
\selectfont\put(67.7969,103.5){\makebox(0,0)[r]{\textcolor[rgb]{0.15,0.15,0.15}{{$0$}}}}
\fontsize{10}{0}
\selectfont\put(67.7969,80.8608){\makebox(0,0)[r]{\textcolor[rgb]{0.15,0.15,0.15}{{$+0.05$}}}}
\fontsize{10}{0}
\selectfont\put(67.7969,58.2222){\makebox(0,0)[r]{\textcolor[rgb]{0.15,0.15,0.15}{{$+0.10$}}}}
\fontsize{10}{0}
\selectfont\put(67.7969,35.5835){\makebox(0,0)[r]{\textcolor[rgb]{0.15,0.15,0.15}{{$+0.15$}}}}
\fontsize{11}{0}
\selectfont\put(246.4,6){\makebox(0,0)[t]{\textcolor[rgb]{0.15,0.15,0.15}{{Radius [AU]}}}}
\fontsize{11}{0}
\selectfont\put(24.7969,103.5){\rotatebox{90}{\makebox(0,0)[b]{\textcolor[rgb]{0.15,0.15,0.15}{{$\theta$ [rad]}}}}}
\fontsize{10}{0}
\selectfont\put(494.448,36.6792){\makebox(0,0)[l]{\textcolor[rgb]{0.15,0.15,0.15}{{400}}}}
\fontsize{10}{0}
\selectfont\put(494.448,57.5742){\makebox(0,0)[l]{\textcolor[rgb]{0.15,0.15,0.15}{{600}}}}
\fontsize{10}{0}
\selectfont\put(494.448,78.4697){\makebox(0,0)[l]{\textcolor[rgb]{0.15,0.15,0.15}{{800}}}}
\fontsize{10}{0}
\selectfont\put(494.448,99.3647){\makebox(0,0)[l]{\textcolor[rgb]{0.15,0.15,0.15}{{1000}}}}
\fontsize{10}{0}
\selectfont\put(494.448,120.26){\makebox(0,0)[l]{\textcolor[rgb]{0.15,0.15,0.15}{{1200}}}}
\fontsize{10}{0}
\selectfont\put(494.448,141.156){\makebox(0,0)[l]{\textcolor[rgb]{0.15,0.15,0.15}{{1400}}}}
\fontsize{10}{0}
\selectfont\put(494.448,162.051){\makebox(0,0)[l]{\textcolor[rgb]{0.15,0.15,0.15}{{1600}}}}
\fontsize{10}{0}
\selectfont\put(494.448,182.946){\makebox(0,0)[l]{\textcolor[rgb]{0.15,0.15,0.15}{{1800}}}}
\end{picture}

%% file: dustdensTC.tex
\setlength{\unitlength}{1pt}
\begin{picture}(0,0)
\includegraphics{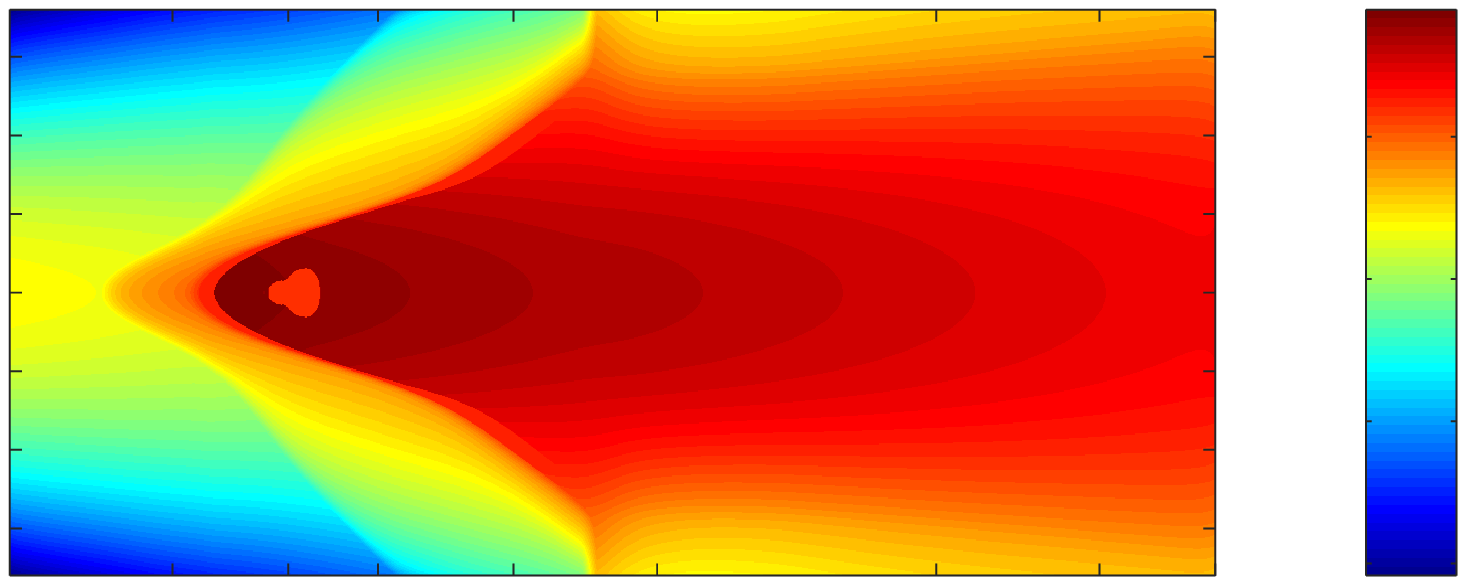}
\end{picture}%
\begin{picture}(560,205)(0,0)
\fontsize{10}{0}
\selectfont\put(446.092,103.5){\rotatebox{90}{\makebox(0,0){\textcolor[rgb]{0,0,0}{{log($\rho_{\mathrm{dust}}$) [g/cm$^3]$}}}}}
\fontsize{10}{0}
\selectfont\put(119.155,17){\makebox(0,0)[t]{\textcolor[rgb]{0.15,0.15,0.15}{{0.3}}}}
\fontsize{10}{0}
\selectfont\put(152.558,17){\makebox(0,0)[t]{\textcolor[rgb]{0.15,0.15,0.15}{{0.4}}}}
\fontsize{10}{0}
\selectfont\put(178.466,17){\makebox(0,0)[t]{\textcolor[rgb]{0.15,0.15,0.15}{{0.5}}}}
\fontsize{10}{0}
\selectfont\put(217.534,17){\makebox(0,0)[t]{\textcolor[rgb]{0.15,0.15,0.15}{{0.7}}}}
\fontsize{10}{0}
\selectfont\put(258.947,17){\makebox(0,0)[t]{\textcolor[rgb]{0.15,0.15,0.15}{{1.0}}}}
\fontsize{10}{0}
\selectfont\put(339.427,17){\makebox(0,0)[t]{\textcolor[rgb]{0.15,0.15,0.15}{{2.0}}}}
\fontsize{10}{0}
\selectfont\put(386.505,17){\makebox(0,0)[t]{\textcolor[rgb]{0.15,0.15,0.15}{{3.0}}}}
\fontsize{10}{0}
\selectfont\put(419.908,17){\makebox(0,0)[t]{\textcolor[rgb]{0.15,0.15,0.15}{{4.0}}}}
\fontsize{10}{0}
\selectfont\put(67.7969,171.417){\makebox(0,0)[r]{\textcolor[rgb]{0.15,0.15,0.15}{{$-0.15$}}}}
\fontsize{10}{0}
\selectfont\put(67.7969,148.778){\makebox(0,0)[r]{\textcolor[rgb]{0.15,0.15,0.15}{{$-0.10$}}}}
\fontsize{10}{0}
\selectfont\put(67.7969,126.139){\makebox(0,0)[r]{\textcolor[rgb]{0.15,0.15,0.15}{{$-0.05$}}}}
\fontsize{10}{0}
\selectfont\put(67.7969,103.5){\makebox(0,0)[r]{\textcolor[rgb]{0.15,0.15,0.15}{{$0$}}}}
\fontsize{10}{0}
\selectfont\put(67.7969,80.8608){\makebox(0,0)[r]{\textcolor[rgb]{0.15,0.15,0.15}{{$+0.05$}}}}
\fontsize{10}{0}
\selectfont\put(67.7969,58.2222){\makebox(0,0)[r]{\textcolor[rgb]{0.15,0.15,0.15}{{$+0.10$}}}}
\fontsize{10}{0}
\selectfont\put(67.7969,35.5835){\makebox(0,0)[r]{\textcolor[rgb]{0.15,0.15,0.15}{{$+0.15$}}}}
\fontsize{11}{0}
\selectfont\put(246.4,6){\makebox(0,0)[t]{\textcolor[rgb]{0.15,0.15,0.15}{{Radius [AU]}}}}
\fontsize{11}{0}
\selectfont\put(24.7969,103.5){\rotatebox{90}{\makebox(0,0)[b]{\textcolor[rgb]{0.15,0.15,0.15}{{$\theta$ [rad]}}}}}
\fontsize{10}{0}
\selectfont\put(494.448,25.5166){\makebox(0,0)[l]{\textcolor[rgb]{0.15,0.15,0.15}{{-30}}}}
\fontsize{10}{0}
\selectfont\put(494.448,66.4692){\makebox(0,0)[l]{\textcolor[rgb]{0.15,0.15,0.15}{{-25}}}}
\fontsize{10}{0}
\selectfont\put(494.448,107.422){\makebox(0,0)[l]{\textcolor[rgb]{0.15,0.15,0.15}{{-20}}}}
\fontsize{10}{0}
\selectfont\put(494.448,148.374){\makebox(0,0)[l]{\textcolor[rgb]{0.15,0.15,0.15}{{-15}}}}
\end{picture}

%% file: dustdensTC2.tex
\setlength{\unitlength}{1pt}
\begin{picture}(0,0)
\includegraphics{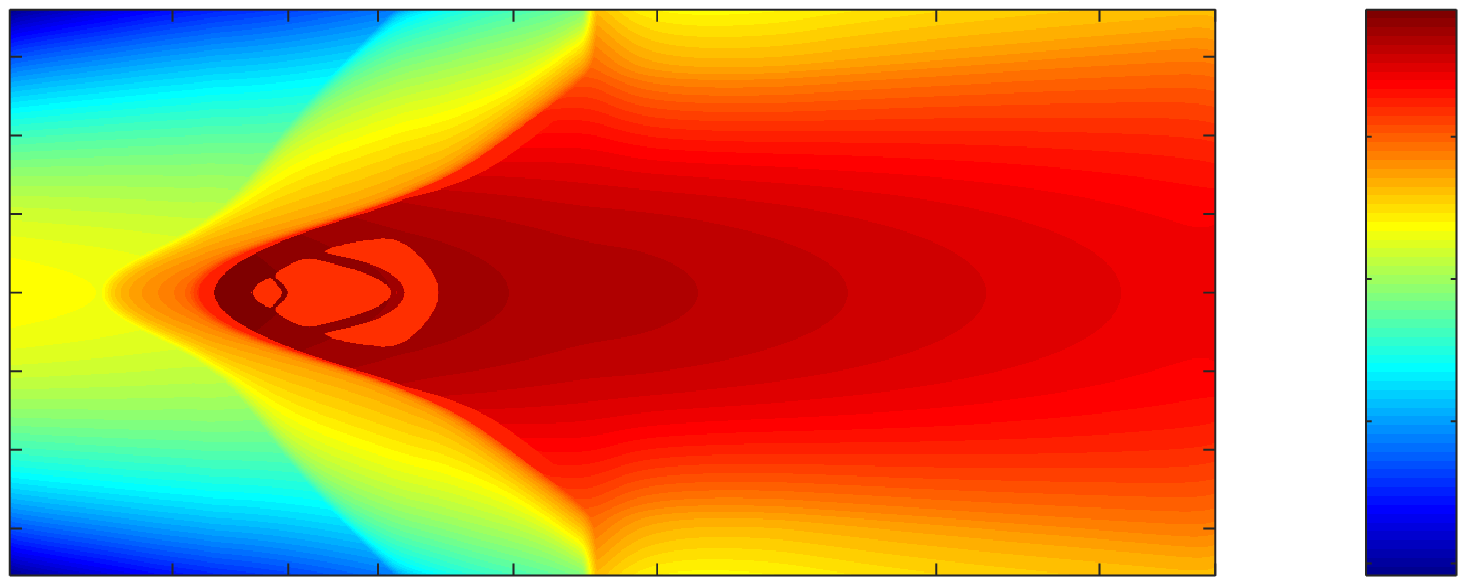}
\end{picture}%
\begin{picture}(560,205)(0,0)
\fontsize{10}{0}
\selectfont\put(446.092,103.5){\rotatebox{90}{\makebox(0,0){\textcolor[rgb]{0,0,0}{{log($\rho_{\mathrm{dust}}$) [g/cm$^3]$}}}}}
\fontsize{10}{0}
\selectfont\put(119.155,17){\makebox(0,0)[t]{\textcolor[rgb]{0.15,0.15,0.15}{{0.3}}}}
\fontsize{10}{0}
\selectfont\put(152.558,17){\makebox(0,0)[t]{\textcolor[rgb]{0.15,0.15,0.15}{{0.4}}}}
\fontsize{10}{0}
\selectfont\put(178.466,17){\makebox(0,0)[t]{\textcolor[rgb]{0.15,0.15,0.15}{{0.5}}}}
\fontsize{10}{0}
\selectfont\put(217.534,17){\makebox(0,0)[t]{\textcolor[rgb]{0.15,0.15,0.15}{{0.7}}}}
\fontsize{10}{0}
\selectfont\put(258.947,17){\makebox(0,0)[t]{\textcolor[rgb]{0.15,0.15,0.15}{{1.0}}}}
\fontsize{10}{0}
\selectfont\put(339.427,17){\makebox(0,0)[t]{\textcolor[rgb]{0.15,0.15,0.15}{{2.0}}}}
\fontsize{10}{0}
\selectfont\put(386.505,17){\makebox(0,0)[t]{\textcolor[rgb]{0.15,0.15,0.15}{{3.0}}}}
\fontsize{10}{0}
\selectfont\put(419.908,17){\makebox(0,0)[t]{\textcolor[rgb]{0.15,0.15,0.15}{{4.0}}}}
\fontsize{10}{0}
\selectfont\put(67.7969,171.417){\makebox(0,0)[r]{\textcolor[rgb]{0.15,0.15,0.15}{{$-0.15$}}}}
\fontsize{10}{0}
\selectfont\put(67.7969,148.778){\makebox(0,0)[r]{\textcolor[rgb]{0.15,0.15,0.15}{{$-0.10$}}}}
\fontsize{10}{0}
\selectfont\put(67.7969,126.139){\makebox(0,0)[r]{\textcolor[rgb]{0.15,0.15,0.15}{{$-0.05$}}}}
\fontsize{10}{0}
\selectfont\put(67.7969,103.5){\makebox(0,0)[r]{\textcolor[rgb]{0.15,0.15,0.15}{{$0$}}}}
\fontsize{10}{0}
\selectfont\put(67.7969,80.8608){\makebox(0,0)[r]{\textcolor[rgb]{0.15,0.15,0.15}{{$+0.05$}}}}
\fontsize{10}{0}
\selectfont\put(67.7969,58.2222){\makebox(0,0)[r]{\textcolor[rgb]{0.15,0.15,0.15}{{$+0.10$}}}}
\fontsize{10}{0}
\selectfont\put(67.7969,35.5835){\makebox(0,0)[r]{\textcolor[rgb]{0.15,0.15,0.15}{{$+0.15$}}}}
\fontsize{11}{0}
\selectfont\put(246.4,6){\makebox(0,0)[t]{\textcolor[rgb]{0.15,0.15,0.15}{{Radius [AU]}}}}
\fontsize{11}{0}
\selectfont\put(24.7969,103.5){\rotatebox{90}{\makebox(0,0)[b]{\textcolor[rgb]{0.15,0.15,0.15}{{$\theta$ [rad]}}}}}
\fontsize{10}{0}
\selectfont\put(494.448,25.5171){\makebox(0,0)[l]{\textcolor[rgb]{0.15,0.15,0.15}{{-30}}}}
\fontsize{10}{0}
\selectfont\put(494.448,66.4692){\makebox(0,0)[l]{\textcolor[rgb]{0.15,0.15,0.15}{{-25}}}}
\fontsize{10}{0}
\selectfont\put(494.448,107.422){\makebox(0,0)[l]{\textcolor[rgb]{0.15,0.15,0.15}{{-20}}}}
\fontsize{10}{0}
\selectfont\put(494.448,148.374){\makebox(0,0)[l]{\textcolor[rgb]{0.15,0.15,0.15}{{-15}}}}
\end{picture}

%% file: Er.tex
\setlength{\unitlength}{1pt}
\begin{picture}(0,0)
\includegraphics{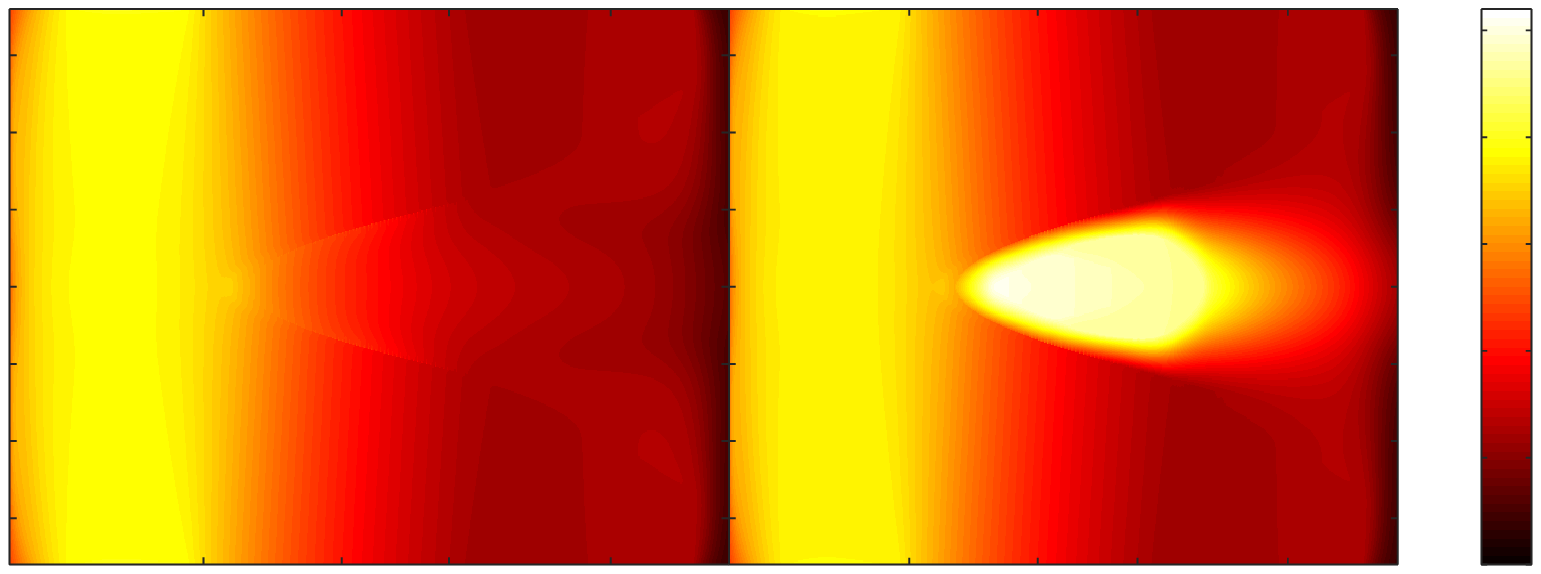}
\end{picture}%
\begin{picture}(560,205)(0,0)
\fontsize{10}{0}
\selectfont\put(111.897,15){\makebox(0,0)[t]{\textcolor[rgb]{0.15,0.15,0.15}{{0.3}}}}
\fontsize{10}{0}
\selectfont\put(151.71,15){\makebox(0,0)[t]{\textcolor[rgb]{0.15,0.15,0.15}{{0.4}}}}
\fontsize{10}{0}
\selectfont\put(182.592,15){\makebox(0,0)[t]{\textcolor[rgb]{0.15,0.15,0.15}{{0.5}}}}
\fontsize{10}{0}
\selectfont\put(229.158,15){\makebox(0,0)[t]{\textcolor[rgb]{0.15,0.15,0.15}{{0.7}}}}
\fontsize{10}{0}
\selectfont\put(50.9951,166.667){\makebox(0,0)[r]{\textcolor[rgb]{0.15,0.15,0.15}{{$-0.15$}}}}
\fontsize{10}{0}
\selectfont\put(50.9951,144.444){\makebox(0,0)[r]{\textcolor[rgb]{0.15,0.15,0.15}{{$-0.10$}}}}
\fontsize{10}{0}
\selectfont\put(50.9951,122.222){\makebox(0,0)[r]{\textcolor[rgb]{0.15,0.15,0.15}{{$-0.05$}}}}
\fontsize{10}{0}
\selectfont\put(50.9951,100){\makebox(0,0)[r]{\textcolor[rgb]{0.15,0.15,0.15}{{$0$}}}}
\fontsize{10}{0}
\selectfont\put(50.9951,77.7778){\makebox(0,0)[r]{\textcolor[rgb]{0.15,0.15,0.15}{{$+0.05$}}}}
\fontsize{10}{0}
\selectfont\put(50.9951,55.5557){\makebox(0,0)[r]{\textcolor[rgb]{0.15,0.15,0.15}{{$+0.10$}}}}
\fontsize{10}{0}
\selectfont\put(50.9951,33.3335){\makebox(0,0)[r]{\textcolor[rgb]{0.15,0.15,0.15}{{$+0.15$}}}}
\fontsize{11}{0}
\selectfont\put(159.6,4){\makebox(0,0)[t]{\textcolor[rgb]{0.15,0.15,0.15}{{Radius [AU]}}}}
\fontsize{11}{0}
\selectfont\put(7.99512,100){\rotatebox{90}{\makebox(0,0)[b]{\textcolor[rgb]{0.15,0.15,0.15}{{$\theta$ [rad]}}}}}
\fontsize{10}{0}
\selectfont\put(315.169,15){\makebox(0,0)[t]{\textcolor[rgb]{0.15,0.15,0.15}{{0.3}}}}
\fontsize{10}{0}
\selectfont\put(352.185,15){\makebox(0,0)[t]{\textcolor[rgb]{0.15,0.15,0.15}{{0.4}}}}
\fontsize{10}{0}
\selectfont\put(380.896,15){\makebox(0,0)[t]{\textcolor[rgb]{0.15,0.15,0.15}{{0.5}}}}
\fontsize{10}{0}
\selectfont\put(424.19,15){\makebox(0,0)[t]{\textcolor[rgb]{0.15,0.15,0.15}{{0.7}}}}
\fontsize{11}{0}
\selectfont\put(359.52,4){\makebox(0,0)[t]{\textcolor[rgb]{0.15,0.15,0.15}{{Radius [AU]}}}}
\fontsize{10}{0}
\selectfont\put(499.528,20){\makebox(0,0)[l]{\textcolor[rgb]{0.15,0.15,0.15}{{0.0000}}}}
\fontsize{10}{0}
\selectfont\put(499.528,50.769){\makebox(0,0)[l]{\textcolor[rgb]{0.15,0.15,0.15}{{0.0005}}}}
\fontsize{10}{0}
\selectfont\put(499.528,81.5386){\makebox(0,0)[l]{\textcolor[rgb]{0.15,0.15,0.15}{{0.0010}}}}
\fontsize{10}{0}
\selectfont\put(499.528,112.308){\makebox(0,0)[l]{\textcolor[rgb]{0.15,0.15,0.15}{{0.0015}}}}
\fontsize{10}{0}
\selectfont\put(499.528,143.077){\makebox(0,0)[l]{\textcolor[rgb]{0.15,0.15,0.15}{{0.0020}}}}
\fontsize{10}{0}
\selectfont\put(499.528,173.846){\makebox(0,0)[l]{\textcolor[rgb]{0.15,0.15,0.15}{{0.0025}}}}
\fontsize{11}{0}
\selectfont\put(475.99512,100){\rotatebox{90}{\makebox(0,0)[b]{\textcolor[rgb]{0.15,0.15,0.15}{{Radiation Energy [J/m$^3$]}}}}}
\end{picture}

%% file: radialdens.tex
\setlength{\unitlength}{0.45pt}
\begin{picture}(0,0)
\includegraphics[scale=0.45]{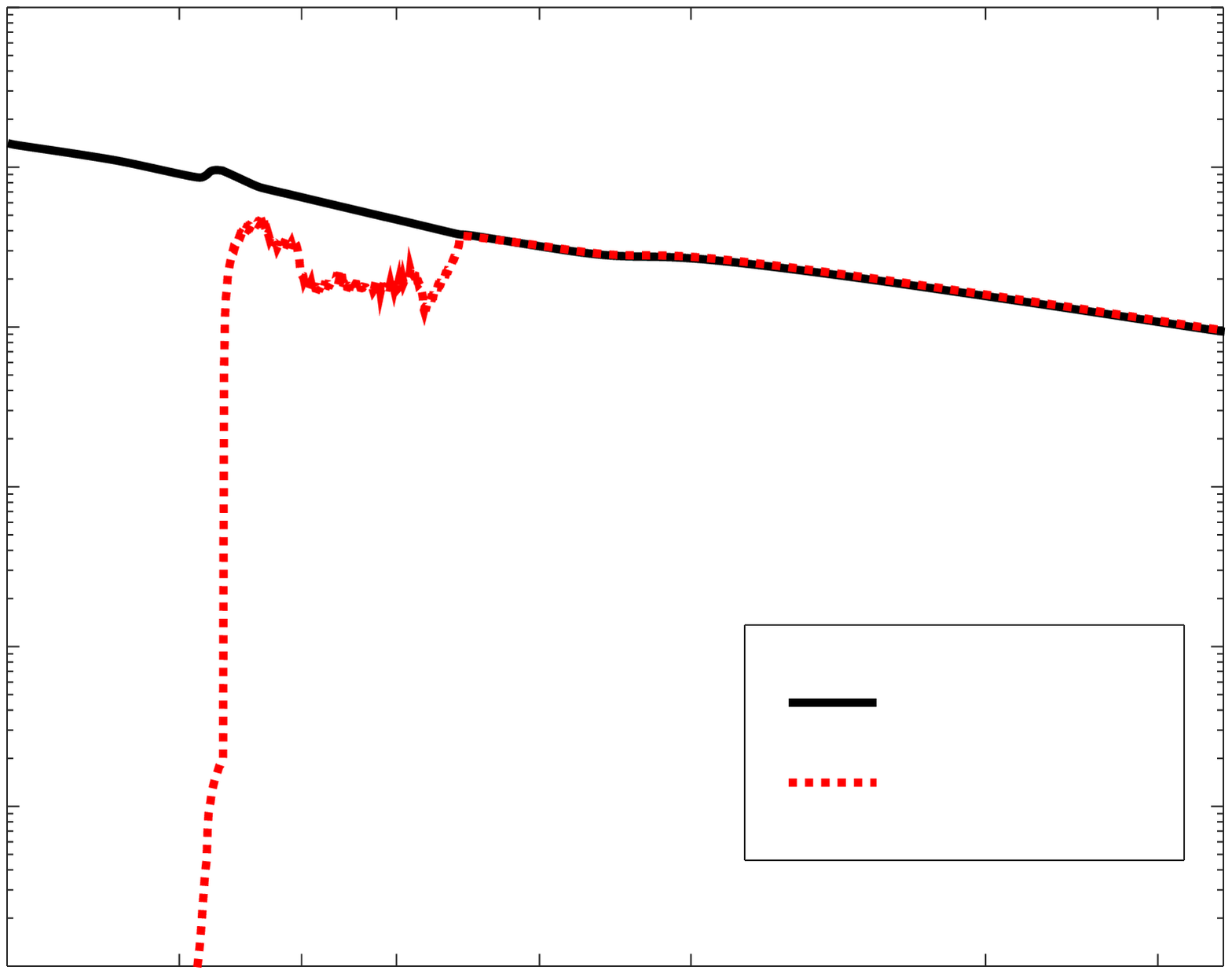}
\end{picture}%
\begin{picture}(576,432)(0,0)
\fontsize{8}{0}
\selectfont\put(74.8799,42.519){\makebox(0,0)[t]{\textcolor[rgb]{0.15,0.15,0.15}{{0.2}}}}
\fontsize{8}{0}
\selectfont\put(138.118,42.519){\makebox(0,0)[t]{\textcolor[rgb]{0.15,0.15,0.15}{{0.3}}}}
\fontsize{8}{0}
\selectfont\put(182.986,42.519){\makebox(0,0)[t]{\textcolor[rgb]{0.15,0.15,0.15}{{0.4}}}}
\fontsize{8}{0}
\selectfont\put(217.788,42.519){\makebox(0,0)[t]{\textcolor[rgb]{0.15,0.15,0.15}{{0.5}}}}
\fontsize{8}{0}
\selectfont\put(270.266,42.519){\makebox(0,0)[t]{\textcolor[rgb]{0.15,0.15,0.15}{{0.7}}}}
\fontsize{8}{0}
\selectfont\put(325.894,42.519){\makebox(0,0)[t]{\textcolor[rgb]{0.15,0.15,0.15}{{1.0}}}}
\fontsize{8}{0}
\selectfont\put(434,42.519){\makebox(0,0)[t]{\textcolor[rgb]{0.15,0.15,0.15}{{2.0}}}}
\fontsize{8}{0}
\selectfont\put(497.238,42.519){\makebox(0,0)[t]{\textcolor[rgb]{0.15,0.15,0.15}{{3.0}}}}
\fontsize{8}{0}
\selectfont\put(69.8755,47.52){\makebox(0,0)[r]{\textcolor[rgb]{0.15,0.15,0.15}{{$10^{-2}$}}}}
\fontsize{8}{0}
\selectfont\put(69.8755,106.2){\makebox(0,0)[r]{\textcolor[rgb]{0.15,0.15,0.15}{{$10^{-1}$}}}}
\fontsize{8}{0}
\selectfont\put(69.8755,164.88){\makebox(0,0)[r]{\textcolor[rgb]{0.15,0.15,0.15}{{$10^{0}$}}}}
\fontsize{8}{0}
\selectfont\put(69.8755,223.56){\makebox(0,0)[r]{\textcolor[rgb]{0.15,0.15,0.15}{{$10^{1}$}}}}
\fontsize{8}{0}
\selectfont\put(69.8755,282.24){\makebox(0,0)[r]{\textcolor[rgb]{0.15,0.15,0.15}{{$10^{2}$}}}}
\fontsize{8}{0}
\selectfont\put(69.8755,340.92){\makebox(0,0)[r]{\textcolor[rgb]{0.15,0.15,0.15}{{$10^{3}$}}}}
\fontsize{8}{0}
\selectfont\put(69.8755,399.6){\makebox(0,0)[r]{\textcolor[rgb]{0.15,0.15,0.15}{{$10^{4}$}}}}
\fontsize{9}{0}
\selectfont\put(298.08,21.519){\makebox(0,0)[t]{\textcolor[rgb]{0.15,0.15,0.15}{{Radius [AU]}}}}
\fontsize{9}{0}
\selectfont\put(34.8755,223.56){\rotatebox{90}{\makebox(0,0)[b]{\textcolor[rgb]{0.15,0.15,0.15}{{$\Sigma$ [g/cm$^2$]}}}}}
\fontsize{8}{0}
\selectfont\put(408.176,144.288){\makebox(0,0)[l]{\textcolor[rgb]{0,0,0}{{$\Sigma_\mathrm{gas}$}}}}
\fontsize{8}{0}
\selectfont\put(408.176,114.912){\makebox(0,0)[l]{\textcolor[rgb]{0,0,0}{{$\Sigma_\mathrm{dust} \cdot 100$}}}}
\end{picture}

%% file: tau.tex
\setlength{\unitlength}{0.45pt}
\begin{picture}(0,0)
\includegraphics[scale=0.45]{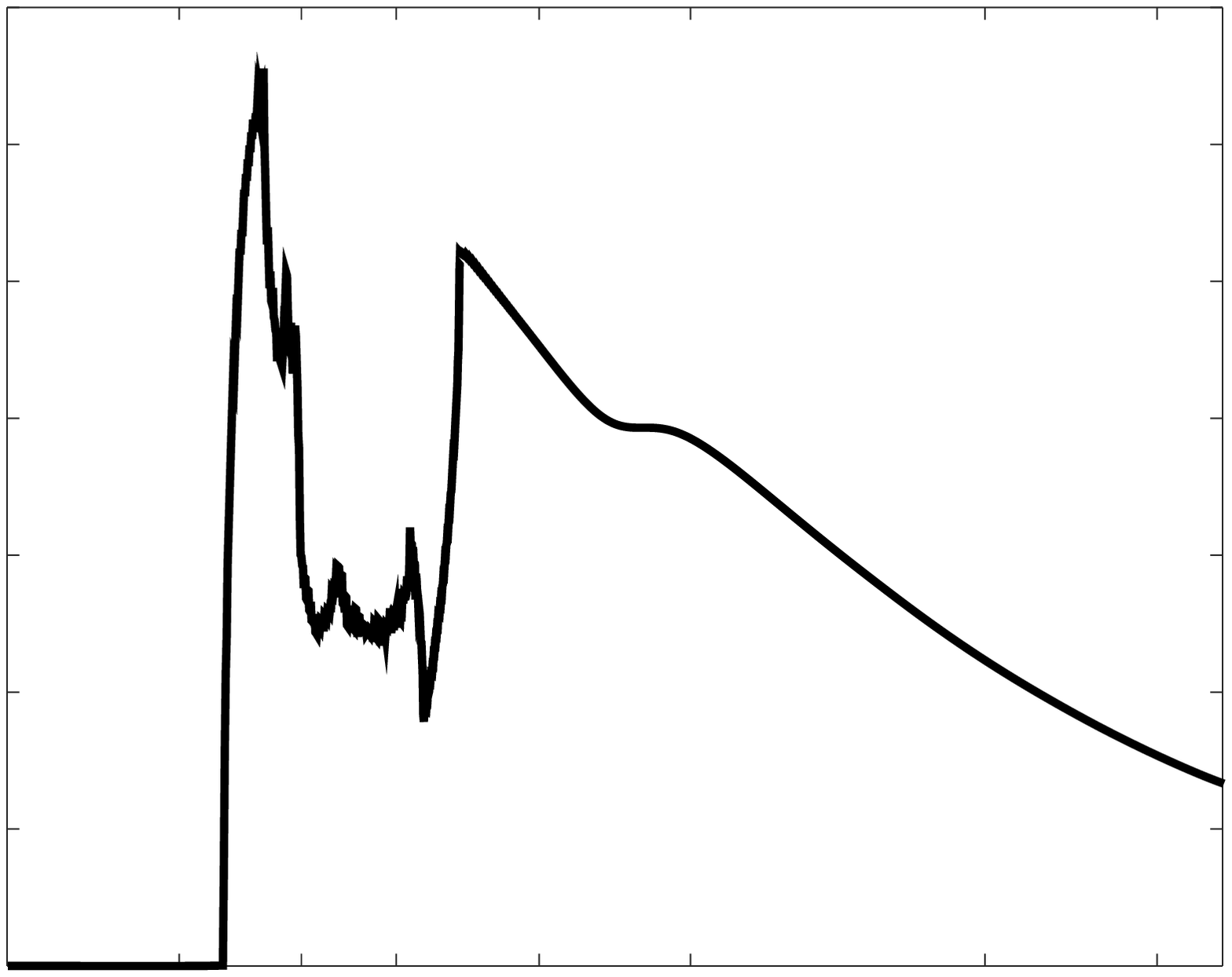}
\end{picture}%
\begin{picture}(576,432)(0,0)
\fontsize{8}{0}
\selectfont\put(74.8799,42.519){\makebox(0,0)[t]{\textcolor[rgb]{0.15,0.15,0.15}{{0.2}}}}
\fontsize{8}{0}
\selectfont\put(138.118,42.519){\makebox(0,0)[t]{\textcolor[rgb]{0.15,0.15,0.15}{{0.3}}}}
\fontsize{8}{0}
\selectfont\put(182.986,42.519){\makebox(0,0)[t]{\textcolor[rgb]{0.15,0.15,0.15}{{0.4}}}}
\fontsize{8}{0}
\selectfont\put(217.788,42.519){\makebox(0,0)[t]{\textcolor[rgb]{0.15,0.15,0.15}{{0.5}}}}
\fontsize{8}{0}
\selectfont\put(270.266,42.519){\makebox(0,0)[t]{\textcolor[rgb]{0.15,0.15,0.15}{{0.7}}}}
\fontsize{8}{0}
\selectfont\put(325.894,42.519){\makebox(0,0)[t]{\textcolor[rgb]{0.15,0.15,0.15}{{1.0}}}}
\fontsize{8}{0}
\selectfont\put(434,42.519){\makebox(0,0)[t]{\textcolor[rgb]{0.15,0.15,0.15}{{2.0}}}}
\fontsize{8}{0}
\selectfont\put(497.238,42.519){\makebox(0,0)[t]{\textcolor[rgb]{0.15,0.15,0.15}{{3.0}}}}
\fontsize{8}{0}
\selectfont\put(69.8755,47.52){\makebox(0,0)[r]{\textcolor[rgb]{0.15,0.15,0.15}{{0}}}}
\fontsize{8}{0}
\selectfont\put(69.8755,97.8169){\makebox(0,0)[r]{\textcolor[rgb]{0.15,0.15,0.15}{{500}}}}
\fontsize{8}{0}
\selectfont\put(69.8755,148.114){\makebox(0,0)[r]{\textcolor[rgb]{0.15,0.15,0.15}{{1000}}}}
\fontsize{8}{0}
\selectfont\put(69.8755,198.412){\makebox(0,0)[r]{\textcolor[rgb]{0.15,0.15,0.15}{{1500}}}}
\fontsize{8}{0}
\selectfont\put(69.8755,248.708){\makebox(0,0)[r]{\textcolor[rgb]{0.15,0.15,0.15}{{2000}}}}
\fontsize{8}{0}
\selectfont\put(69.8755,299.006){\makebox(0,0)[r]{\textcolor[rgb]{0.15,0.15,0.15}{{2500}}}}
\fontsize{8}{0}
\selectfont\put(69.8755,349.303){\makebox(0,0)[r]{\textcolor[rgb]{0.15,0.15,0.15}{{3000}}}}
\fontsize{8}{0}
\selectfont\put(69.8755,399.6){\makebox(0,0)[r]{\textcolor[rgb]{0.15,0.15,0.15}{{3500}}}}
\fontsize{9}{0}
\selectfont\put(298.08,21.519){\makebox(0,0)[t]{\textcolor[rgb]{0.15,0.15,0.15}{{Radius [AU]}}}}
\fontsize{9}{0}
\selectfont\put(21.8755,223.56){\rotatebox{90}{\makebox(0,0)[b]{\textcolor[rgb]{0.15,0.15,0.15}{{$\tau_\mathrm{z}$}}}}}
\end{picture}